\documentclass[a4paper,12pt]{article}
\pdfoutput=1
\usepackage{a4wide}
\usepackage[bookmarks=false,colorlinks]{hyperref}
\usepackage[pdftex]{graphicx}
\usepackage{amsmath,amsfonts,amssymb,mathtools,bm}
\usepackage{rotating}
\usepackage{tabularx}
\usepackage{multicol,multirow}
\usepackage{latexsym}
\usepackage{relsize}
\usepackage[center]{subfigure}
\usepackage{float}
\usepackage{caption}
\usepackage{slashed}
\usepackage{lineno}
\usepackage{cite,mcite}
\usepackage{appendix}
\usepackage{url}
\usepackage{soul}
\usepackage{pifont}
\hypersetup{urlcolor=blue, citecolor=blue}
\usepackage{colortbl}
\usepackage{hhline}

\date{\today}
 
\normalsize

\def\Bbar{\overline{B}}

\def\cbar{\overline{c}}

\def\taubar{\overline{\tau}}

\def\nubar{{\overline{\nu}}}

\def\Heff{\mathcal{H}_{\rm eff}}
\def\L{\mathcal{L}}

\def\A{\mathcal{A}}
\def\O{\mathcal{O}}
\def\B{\mathcal{B}}
\def\Re{\mathcal{R}e}

\def\Dst{{D^*}}
\def\DDst{{D^{(*)}}}

\def\SM{{\rm SM}}
\def\GeV{{\rm GeV}}
\def\AFB{{\A_{\rm FB}}}

\usepackage{relsize}
\def\Babar{{\mbox{\slshape B\kern-0.1em{\smaller A}\kern-0.1em B\kern-0.1em{\smaller A\kern-0.2em R}}}}
\def\markR{{\tiny\color{blue}\bm\blacksquare}}
\def\markRq{{\color{red}\bm\circ}}
\def\MarkRq{{\color{red}\bm\circledcirc}}

\allowdisplaybreaks

\begin{document}

\thispagestyle{empty} 
\rightline{KEK-TH-1779}
\rightline{OU-HET 838}
\rightline{CTPU-14-14}
\vspace{1.5cm} 
{\Large
\begin{center}
   {\bf Probing New Physics with $\bm{q^2}$ distributions in $\bm{\Bbar \to D^{(*)} \tau\nubar}$}
\end{center}
}
\vspace{0.3cm}

\begin{center}
   {\sc Yasuhito Sakaki} \\
   \vspace{5mm}
   {\small\emph{Theory Group, IPNS, KEK, Tsukuba, Ibaraki 305-0801, Japan}} \\
   \vspace{5mm}
   {\sc Minoru Tanaka, Andrey Tayduganov\,\footnote{Present affiliation: CPPM, CNRS/IN2P3 and Aix-Marseille Universit\'e.}} \\
   \vspace{5mm}
   {\small\emph{Department of Physics, Graduate School of Science, Osaka University,}} \\
   {\small\emph{Toyonaka, Osaka 560-0043, Japan}} \\
   \vspace{5mm}
   {\sc Ryoutaro Watanabe} \\
   \vspace{5mm}
   {\small\emph{Center for Theoretical Physics of the Universe, Institute for Basic Science (IBS),}} \\
   {\small\emph{Daejeon 305-811, Republic of Korea}}
\end{center}

\vskip1cm
\begin{center}
   \small{\bf Abstract}
\end{center}

\vspace{3mm}
Recent experimental results for the ratios of the branching fractions of the decays $\Bbar \to \DDst\tau\nubar$ and $\Bbar \to \DDst\mu\nubar$ came as a surprise and lead to a discussion of possibility of testing New Physics beyond the Standard Model through these modes. We show that these decay channels can provide us with good constraints on New Physics and several New Physics cases are favored by the present experimental data. In order to discriminate various New Physics scenarios, we examine the $q^2$ distributions and estimate the sensitivity of this potential measurement at the SuperKEKB/Belle~II experiment.


\newpage

\section{Introduction}\label{sec:intro}

Recently, the \Babar~ and Belle collaborations observed excess of exclusive semitauonic decays of $B$ meson, $\Bbar \to D\tau\nubar$ and $\Bbar \to \Dst\tau\nubar$. In order to test the lepton universality with less theoretical uncertainty, the ratios of the branching fractions are introduced as observables,
\begin{equation}
   R(\DDst) \equiv {\B(\Bbar \to \DDst\tau\nubar) \over \B(\Bbar \to \DDst\ell\nubar)} \,,
\end{equation}
where $\ell$ denotes $e$ or $\mu$.
Combining the \Babar~\cite{Lees:2012xj,Lees:2013uzd} and Belle \cite{Matyja:2007kt,*Adachi:2009qg,*Bozek:2010xy} results for $R(D)$ and $R(\Dst)$, we obtain
\begin{equation}
   R(D)=0.421\pm 0.058 \,, \quad R(D^*)=0.337\pm 0.025 \,,
   \label{Eq:combined}
\end{equation}
with the correlation to be $-0.19$.
Comparing it to the Standard Model (SM) predictions,
\begin{equation}
   R(D)^\SM = 0.305\pm 0.012 \,, \quad R(\Dst)^\SM = 0.252\pm 0.004 \,,
   \label{Eq:SMP}
\end{equation}
we find a discrepancy of $3.5\sigma$.

From the theoretical point of view, the two-Higgs-doublet model of type II (2HDM-II), which is the Higgs sector of the minimal supersymmetric Standard Model, has been studied well in the literature as a candidate of New Physics (NP) beyond the SM that significantly affects the semitauonic $B$ decays \cite{Hou:1992sy,*Tanaka:1994ay,*Nierste:2008qe,*Kamenik:2008tj,*Tanaka:2010se}. Using the results of these theoretical works and the experimental data, the \Babar~ collaboration shows that the 2HDM-II is excluded at 99.8\% confidence level (C.L.)~\cite{Lees:2012xj,Lees:2013uzd}.

This observation has stimulated further theoretical activities for clarifying the origin of the above discrepancy. Possible structures of the relevant four-fermion interaction are identified and NP models (other than 2HDM-II) that could induce such structures are proposed in the literature \cite{Tanaka:2012nw,Sakaki:2013bfa,Crivellin:2012ye,*Ko:2012sv,*Sakaki:2012ft,*Celis:2012dk,*Celis:2013jha,*Fajfer:2012vx,*Fajfer:2012jt,*Becirevic:2012jf,*Bailey:2012jg,*Datta:2012qk,*Biancofiore:2013ki,*Hagiwara:2014tsa}.

For further tests and discrimination of the allowed NP models, in Ref.~\cite{Sakaki:2013bfa} we examined various correlations among the $\tau$ forward-backward asymmetries, the $\tau$ polarizations and the $\Dst$ longitudinal polarization in some favorable cases. However, one has to note that the measurement of $\AFB$, $P_\tau$ and $P_\Dst$ is a challenging (but feasible) experimental task due to the missing energy/momentum of neutrinos in $\tau$ decay reconstruction and the tiny phase space in $\Dst \to D\pi$ decay. Therefore, besides the above integrated quantities $R(\DDst)$, in this work we study the possibility of discriminating various NP scenarios using the ratios of differential branching fractions that could be also sensitive to NP.

In Section \ref{sec:Heff}, we introduce the effective Hamiltonian, describing the $\Bbar\to \DDst\tau\nubar$ decays, and put constraints on the NP Wilson coefficients. In Section \ref{sec:distributions}, we study the NP effects in the $q^2\equiv(p_B-p_\DDst)^2$ distributions of the differential branching fractions and introduce new quantities $R_\DDst(q^2)$. In Section \ref{sec:BelleII}, we demonstrate that $R_\DDst(q^2)$ could be particularly helpful in discriminating between various NP operators. We also examine the sensitivity of the future measurement at the SuperKEKB/Belle~II experiment.

\section{Effective Hamiltonian and New Physics constraints}\label{sec:Heff}

Assuming the neutrinos to be left-handed, we introduce the most general effective Hamiltonian that contains all possible four-fermion operators of the lowest dimension for the $b \to c\tau\nubar_\tau$ transition\,\footnote{In our work, we assume that couplings of NP particles to light leptons are significantly suppressed (as in the 2HDM-II) and NP effects can be observed only in the tauonic decay modes.},
\begin{equation}
   \Heff = {4G_F \over \sqrt2} V_{cb}\left[ (1 + C_{V_1})\O_{V_1} + C_{V_2}\O_{V_2} + C_{S_1}\O_{S_1} + C_{S_2}\O_{S_2} + C_T\O_T \right] \,,
      \label{eq:Heff}
\end{equation}
where the operator basis is defined as
\begin{equation}
   \begin{split}
      \O_{V_1} =& (\cbar_L \gamma^\mu b_L)(\taubar_L \gamma_\mu \nu_{L}) \,, \\
      \O_{V_2} =& (\cbar_R \gamma^\mu b_R)(\taubar_L \gamma_\mu \nu_{L}) \,, \\
      \O_{S_1} =& (\cbar_L b_R)(\taubar_R \nu_{L}) \,, \\
      \O_{S_2} =& (\cbar_R b_L)(\taubar_R \nu_{L}) \,, \\
      \O_T =& (\cbar_R \sigma^{\mu\nu} b_L)(\taubar_R \sigma_{\mu\nu} \nu_{L}) \,,
   \end{split}
   \label{eq:operators}
\end{equation}
and the neutrino flavor is assumed to be identical to the SM one. In the SM, the Wilson coefficients are set to zero, $C_X^{\rm (SM)}=0$ ($X=V_{1,2},\,S_{1,2},\,T$). In Ref.~\cite{Tanaka:2012nw}, all five generic operators were studied. It was demonstrated that vector $\O_{V_{1,2}}$, scalar $\O_{S_2}$ and tensor $\O_T$ operators can reasonably explain the current data, and the scalar $\O_{S_1}$ is unlikely.

\begin{figure}[t!]\centering
   \includegraphics[width=0.32\linewidth]{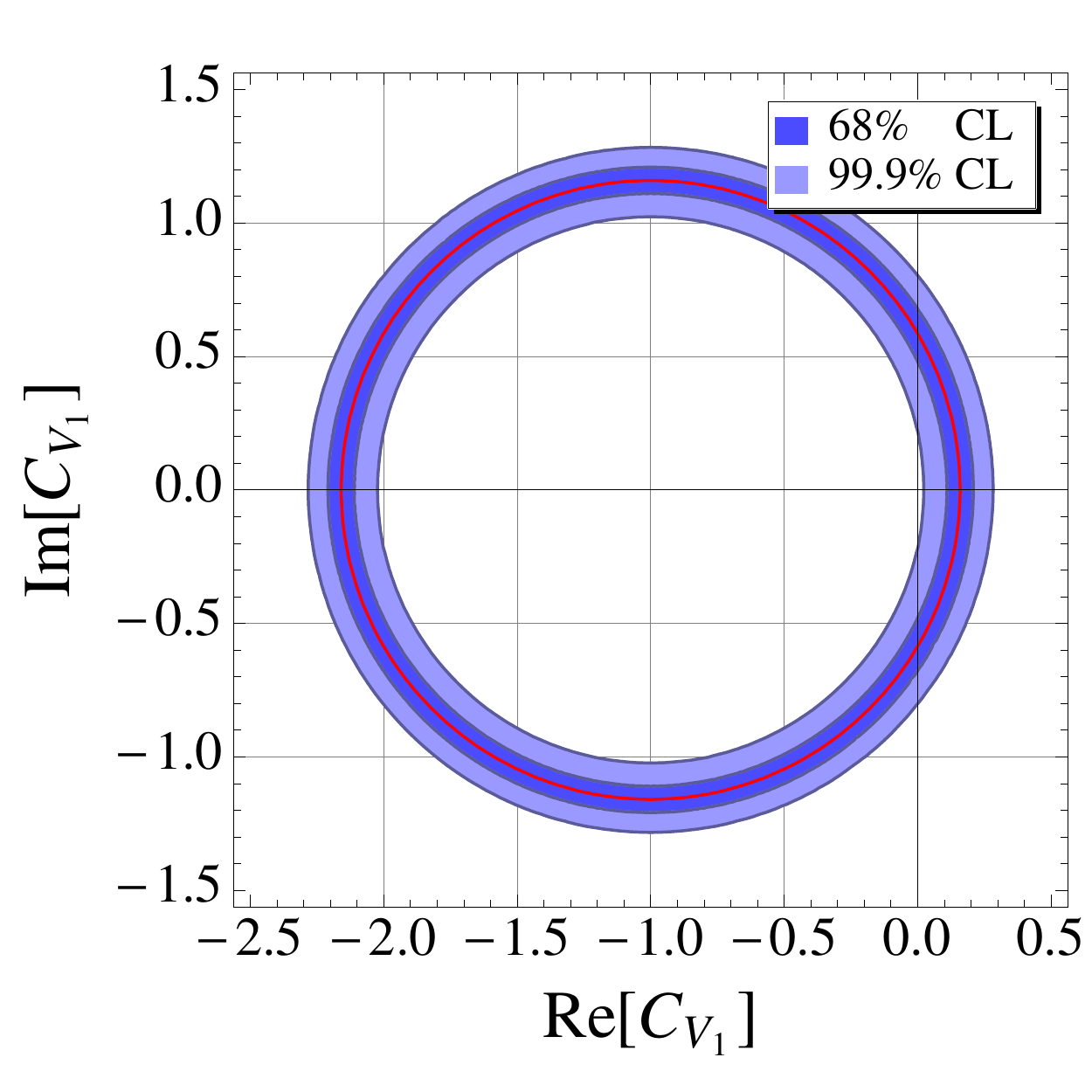}
   \includegraphics[width=0.32\linewidth]{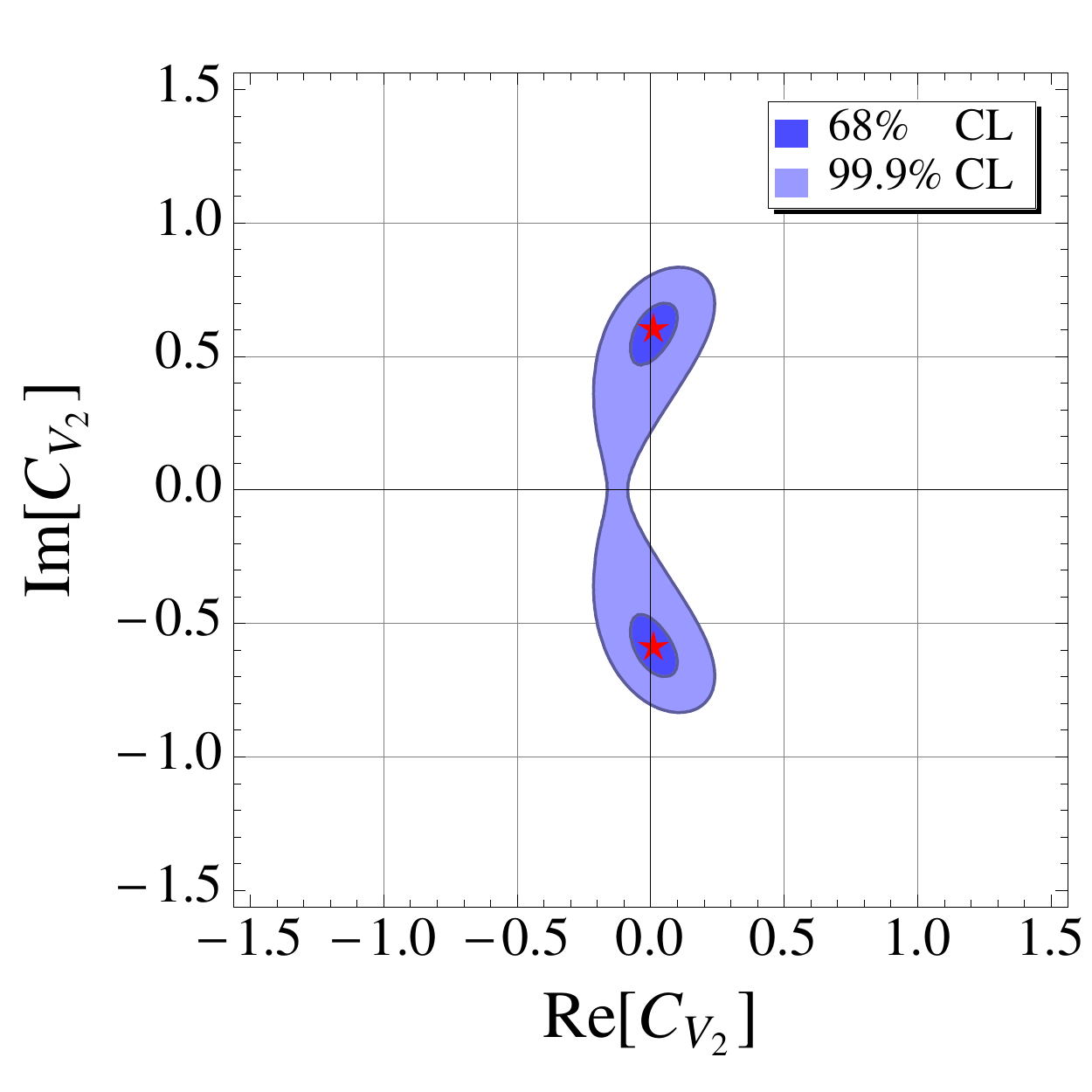}
   \includegraphics[width=0.32\linewidth]{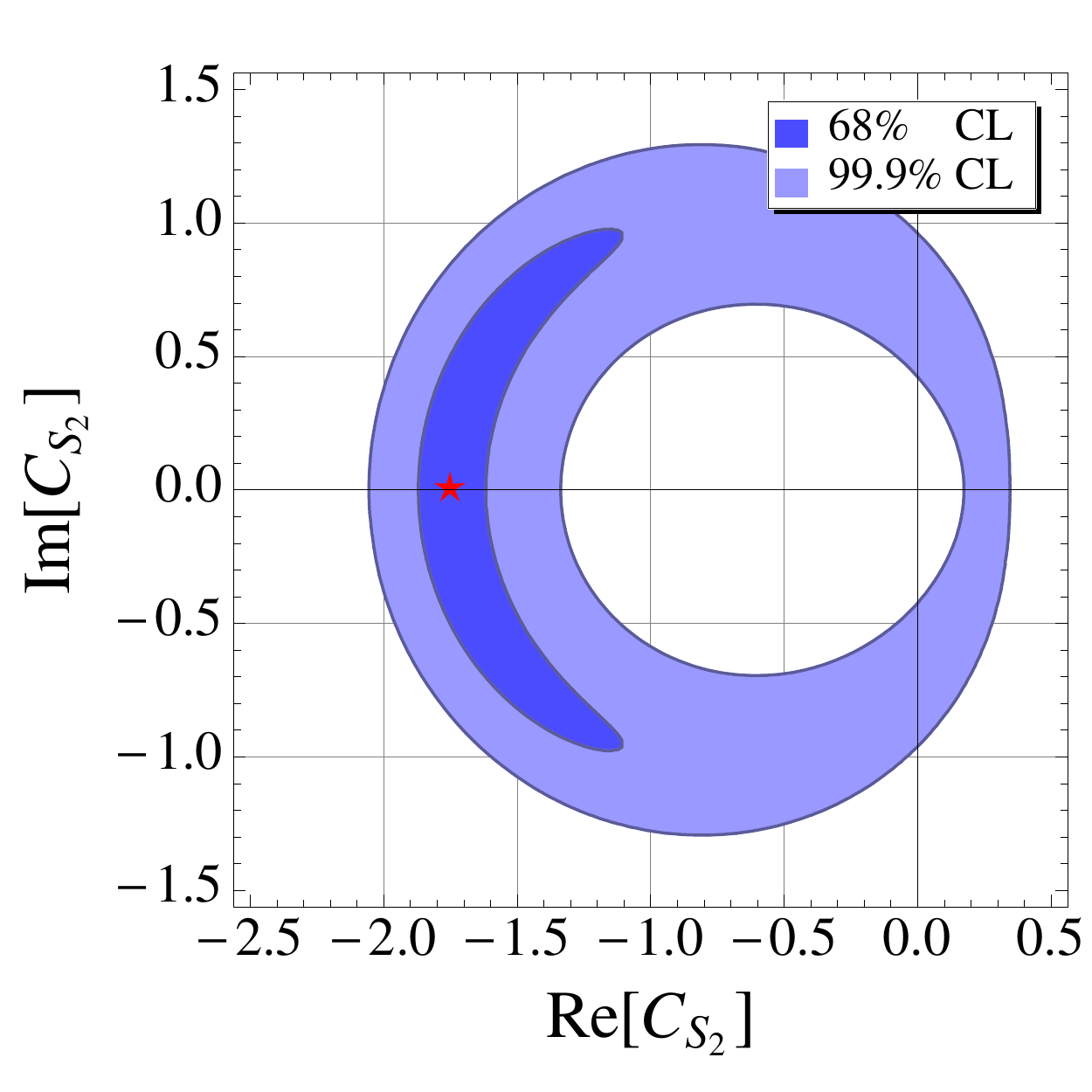}
   \put(-410,125){(a)}
   \put(-260,125){(b)}
   \put(-110,125){(c)}

   \includegraphics[width=0.32\linewidth]{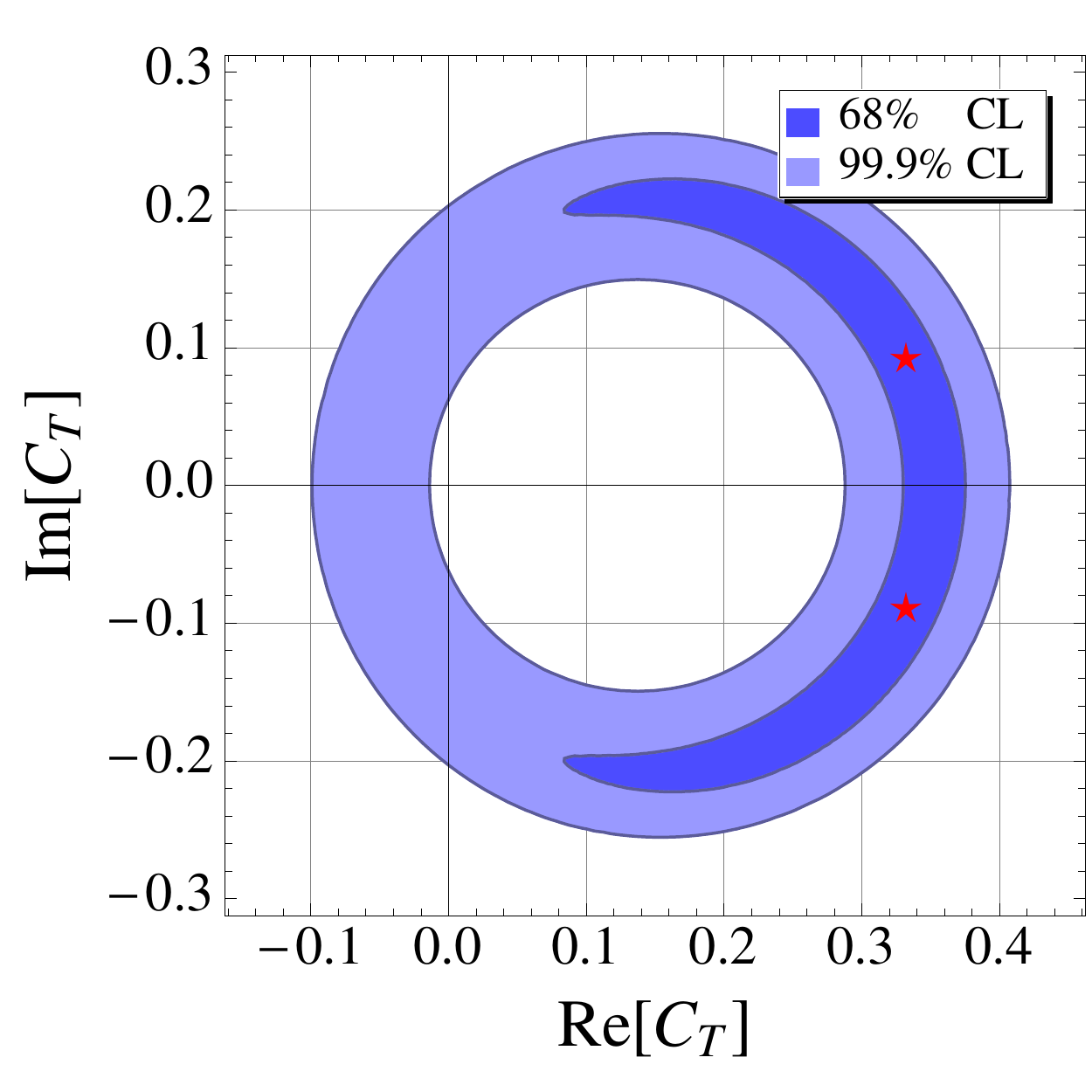}
   \includegraphics[width=0.32\linewidth]{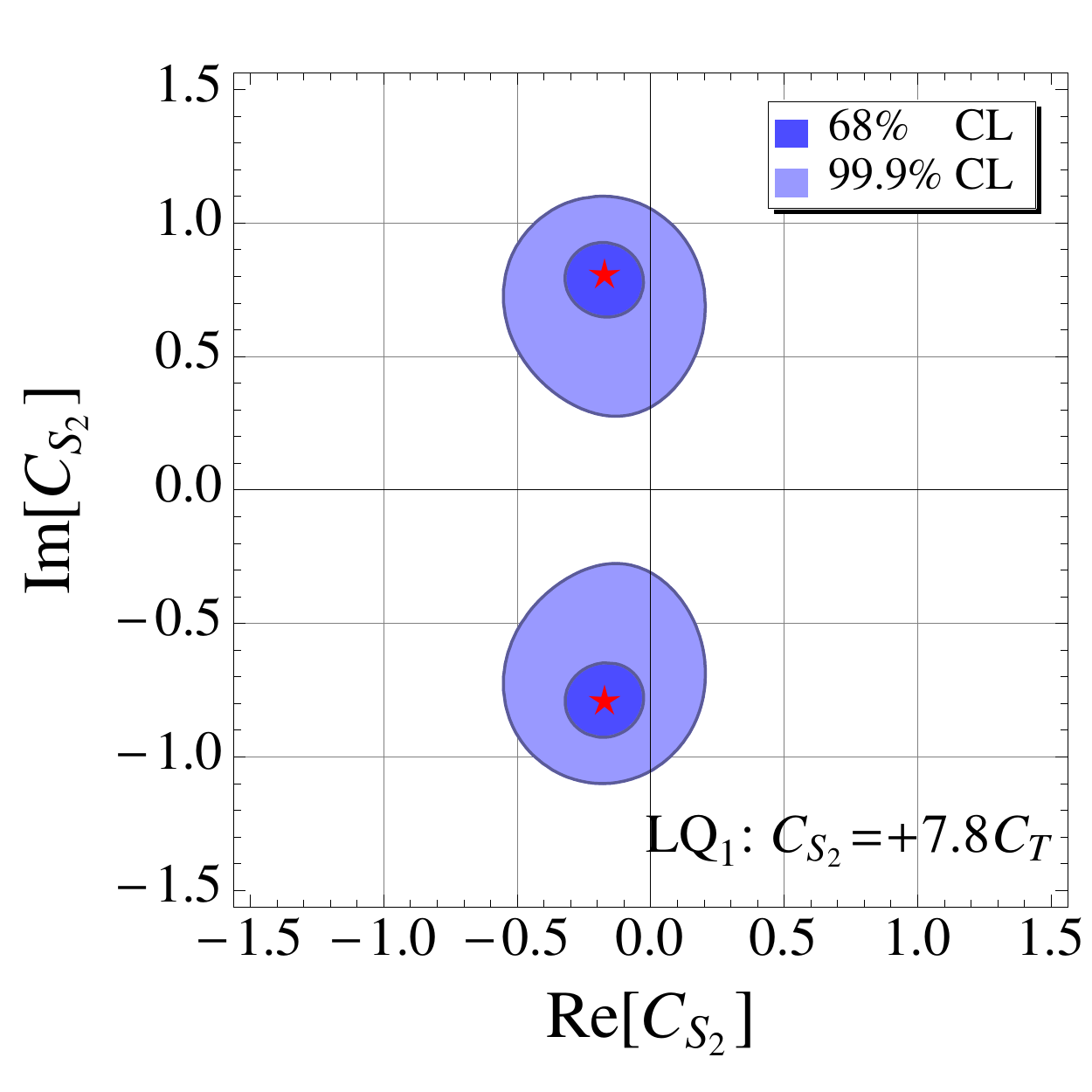}
   \includegraphics[width=0.32\linewidth]{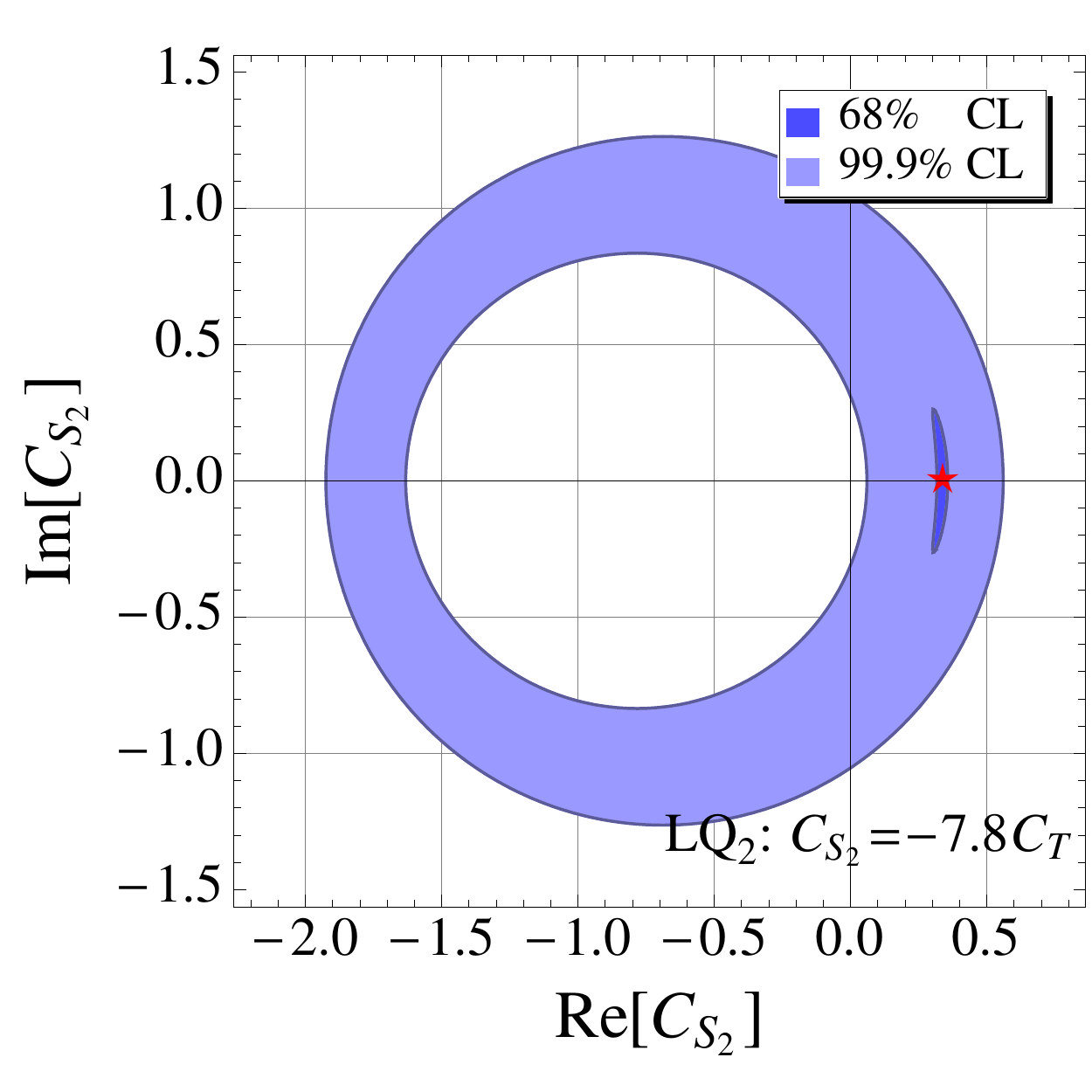}
   \put(-410,125){(d)}
   \put(-260,125){(e)}
   \put(-110,125){(f)}
   \caption{\footnotesize Constraints on the Wilson coefficients at the $m_b$ scale. The constraints are obtained from the $\chi^2$ fit of the measured $R(D)$ and $R(\Dst)$. The stars represent the optimal fitted values giving the smallest $\chi^2$.}
   \label{pic:CX}
\end{figure}

In Fig.~\ref{pic:CX} the allowed regions for complex NP Wilson coefficients at the bottom quark mass scale are shown, obtained from the $\chi^2$ fit of the current \Babar~ and Belle measurements of $R(D)$ and $R(\Dst)$ in Eq.~\eqref{Eq:combined}. We assume the presence of only one NP operator for (a)-(d); and two operators $\O_{S_2}$ and $\O_T$ for (e) and (f), for which the Wilson coefficients are related as $C_{S_2}=\pm7.8C_T$ at the $m_b$ scale\,\footnote{This ratio is obtained from the renormalization group running of the scalar and tensor operators from the leptoquark mass scale of 1~TeV, at which one finds $C_{S_2}=\pm4C_T$, down to the $m_b$ scale \cite{Dorsner:2013tla}.} as written in the figure. These NP types of scalar and tensor exist in leptoquark models \cite{Sakaki:2013bfa,Tanaka:2012nw,Lee:2001nw,Dorsner:2013tla}. The star corresponds to the best fitted values giving the smallest $\chi^2$ value. We note that, since $\B=|1+C_{V_1}|^2\B^\SM$, the best fitted value for $C_{V_1}$ is degenerate and represented by the red circle on the left-top panel of Fig.~\ref{pic:CX}. One can see that Wilson coefficients of $O(1)$ are sufficient to explain the observed discrepancy in $R(D)$ and $R(\Dst)$.

Minimizing $\chi^2$ and finding the optimal NP Wilson coefficients, in the following sections we study various scenarios as benchmarks:
\begin{itemize}
   \item SM : $C_X = 0$ \,,
   \item $V_1$ : $C_{V_1}=0.16$, $C_{X\neq V_1} = 0$ \,,
   \item $V_2$ : $C_{V_2}=0.01\pm0.60i$, $C_{X\neq V_2} = 0$ \,,
   \item $S_2$ : $C_{S_2} = -1.75$, $C_{X\neq S_2} = 0$ \,,
   \item $T$ : $C_T = 0.33\pm0.09i$, $C_{X\neq T} = 0$ \,,
   \item LQ$_1$ scenario: $C_{S_2}=7.8C_T=-0.17\pm0.80i$, $C_{X\neq S_2,T} = 0$ \,,
   \item LQ$_2$ scenario: $C_{S_2}=-7.8C_T=0.34$, $C_{X\neq S_2,T} = 0$ \,.
\end{itemize}

\section{New Physics effects in the $\bm{q^2}$ distributions}\label{sec:distributions}

Using the effective Hamiltonian in Eq.~\eqref{eq:Heff} and calculating the helicity amplitudes (for the details see Ref. \cite{Tanaka:2012nw}), one finds the differential decay rates as follows \cite{Sakaki:2013bfa} :

\begin{equation}
   \begin{split}
      {d\Gamma(\Bbar \to D \tau\nubar_\tau) \over dq^2} =& {G_F^2 |V_{cb}|^2 \over 192\pi^3 m_B^3} q^2 \sqrt{\lambda_D(q^2)} \left( 1 - {m_\tau^2 \over q^2} \right)^2 \times\biggl\{ \biggr. \\
                                                         & |1 + C_{V_1} + C_{V_2}|^2 \left[ \left( 1 + {m_\tau^2 \over2q^2} \right) H_{V,0}^{s\,2} + {3 \over 2}{m_\tau^2 \over q^2} \, H_{V,t}^{s\,2} \right] \\
                                                         &+ {3 \over 2} |C_{S_1} + C_{S_2}|^2 \, H_S^{s\,2} + 8|C_T|^2 \left( 1+ {2m_\tau^2 \over q^2} \right) \, H_T^{s\,2} \\
                                                         &+ 3\Re[ ( 1 + C_{V_1} + C_{V_2} ) (C_{S_1}^{*} + C_{S_2}^{*} ) ] {m_\tau \over \sqrt{q^2}} \, H_S^s H_{V,t}^s \\
                                                         &- 12\Re[ ( 1 + C_{V_1} + C_{V_2} ) C_T^{*} ] {m_\tau \over \sqrt{q^2}} \, H_T^s H_{V,0}^s \biggl.\biggr\} \,,
   \end{split}
\end{equation}
and

\begin{equation}
   \begin{split}
      & {d\Gamma(\Bbar \to \Dst \tau\nubar_\tau) \over dq^2} = {G_F^2 |V_{cb}|^2 \over 192\pi^3 m_B^3} q^2 \sqrt{\lambda_\Dst(q^2)} \left( 1 - {m_\tau^2 \over q^2} \right)^2 \times\biggl\{ \biggr. \\
      & \quad\quad\quad\quad\quad ( |1 + C_{V_1}|^2 + |C_{V_2}|^2 ) \left[ \left( 1 + {m_\tau^2 \over2q^2} \right) \left( H_{V,+}^2 + H_{V,-}^2 + H_{V,0}^2 \right) + {3 \over 2}{m_\tau^2 \over q^2} \, H_{V,t}^2 \right] \\
      & \quad\quad\quad\quad\quad - 2\Re[(1 + C_{V_1}) C_{V_2}^{*}] \left[ \left( 1 + {m_\tau^2 \over 2q^2} \right) \left( H_{V,0}^2 + 2 H_{V,+} H_{V,-} \right) + {3 \over 2}{m_\tau^2 \over q^2} \, H_{V,t}^2 \right] \\
      & \quad\quad\quad\quad\quad + {3 \over 2} |C_{S_1} - C_{S_2}|^2 \, H_S^2 + 8|C_T|^2 \left( 1+ {2m_\tau^2 \over q^2} \right) \left( H_{T,+}^2 + H_{T,-}^2 + H_{T,0}^2  \right) \\
      & \quad\quad\quad\quad\quad + 3\Re[ (1 + C_{V_1} - C_{V_2} ) (C_{S_1}^{*} - C_{S_2}^{*} ) ] {m_\tau \over \sqrt{q^2}} \, H_S H_{V,t} \\
      & \quad\quad\quad\quad\quad - 12\Re[ (1 + C_{V_1}) C_T^{*} ] {m_\tau \over \sqrt{q^2}} \left( H_{T,0} H_{V,0} + H_{T,+} H_{V,+} - H_{T,-} H_{V,-} \right) \\
      & \quad\quad\quad\quad\quad + 12\Re[ C_{V_2} C_T^{*} ] {m_\tau \over \sqrt{q^2}} \left( H_{T,0} H_{V,0} + H_{T,+} H_{V,-} - H_{T,-} H_{V,+} \right) \biggl.\biggr\} \,,
   \end{split}
\end{equation}
where $\lambda_{D^{(*)}}(q^2) = ((m_B - m_{D^{(*)}})^2 - q^2)((m_B + m_{D^{(*)}})^2 - q^2)$. The SM distributions for the light lepton modes can be easily obtained by setting $C_X=0$ and $m_\tau=0$.

The helicity amplitudes $H$'s are expressed in terms of hadronic $\Bbar \to \DDst$ form factors. In this work we use the Heavy Quark Effective Theory (HQET) form factors \cite{Caprini:1997mu} with parameters extracted from experiments by the \Babar~ and Belle collaborations \cite{Amhis:2012bh}. A detailed description of the matrix elements and form factor parametrization can be found in Ref.~\cite{Sakaki:2013bfa}.

\begin{figure}[t!]\centering
   \includegraphics[width=0.45\linewidth]{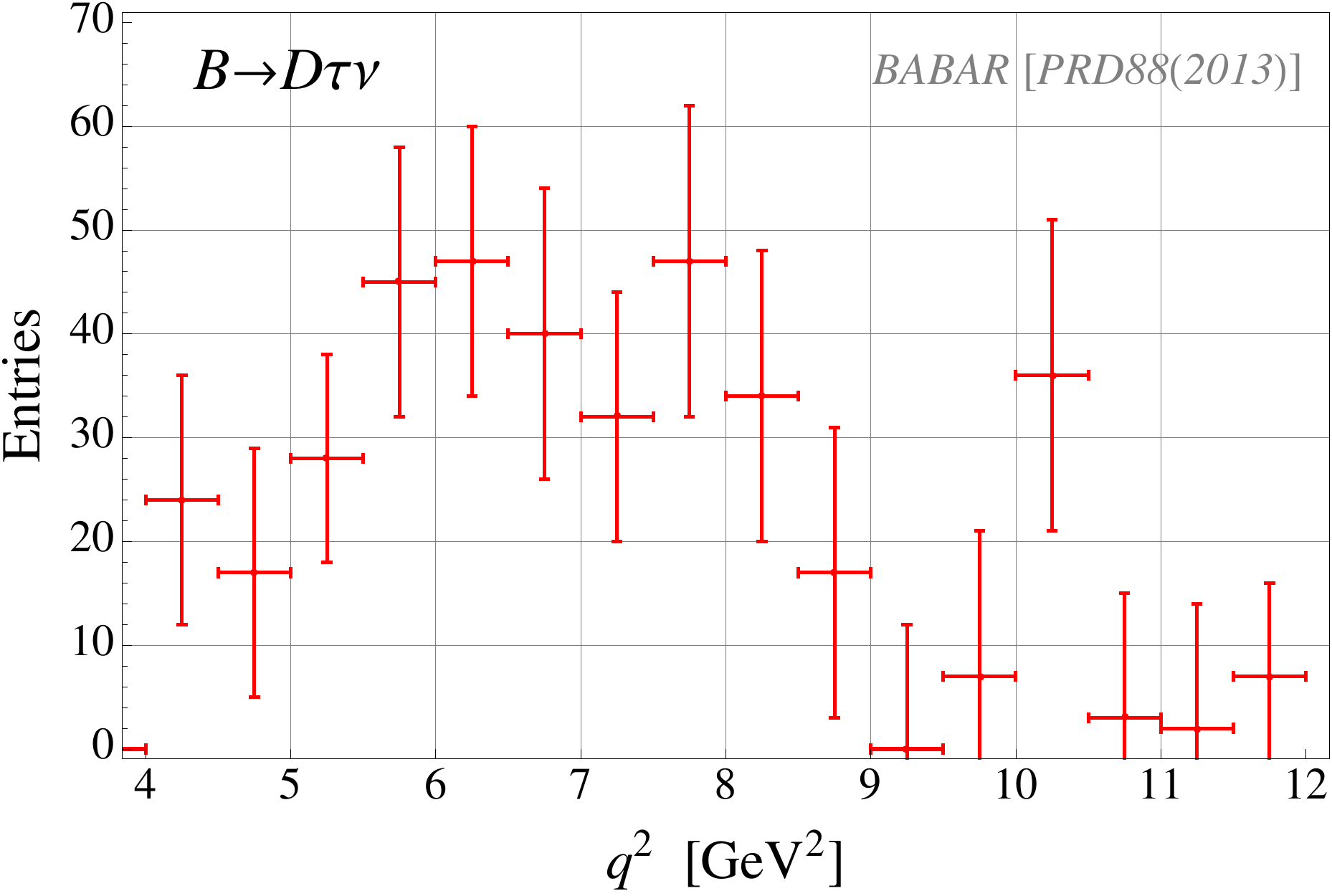} \hspace{3mm}
   \includegraphics[width=0.45\linewidth]{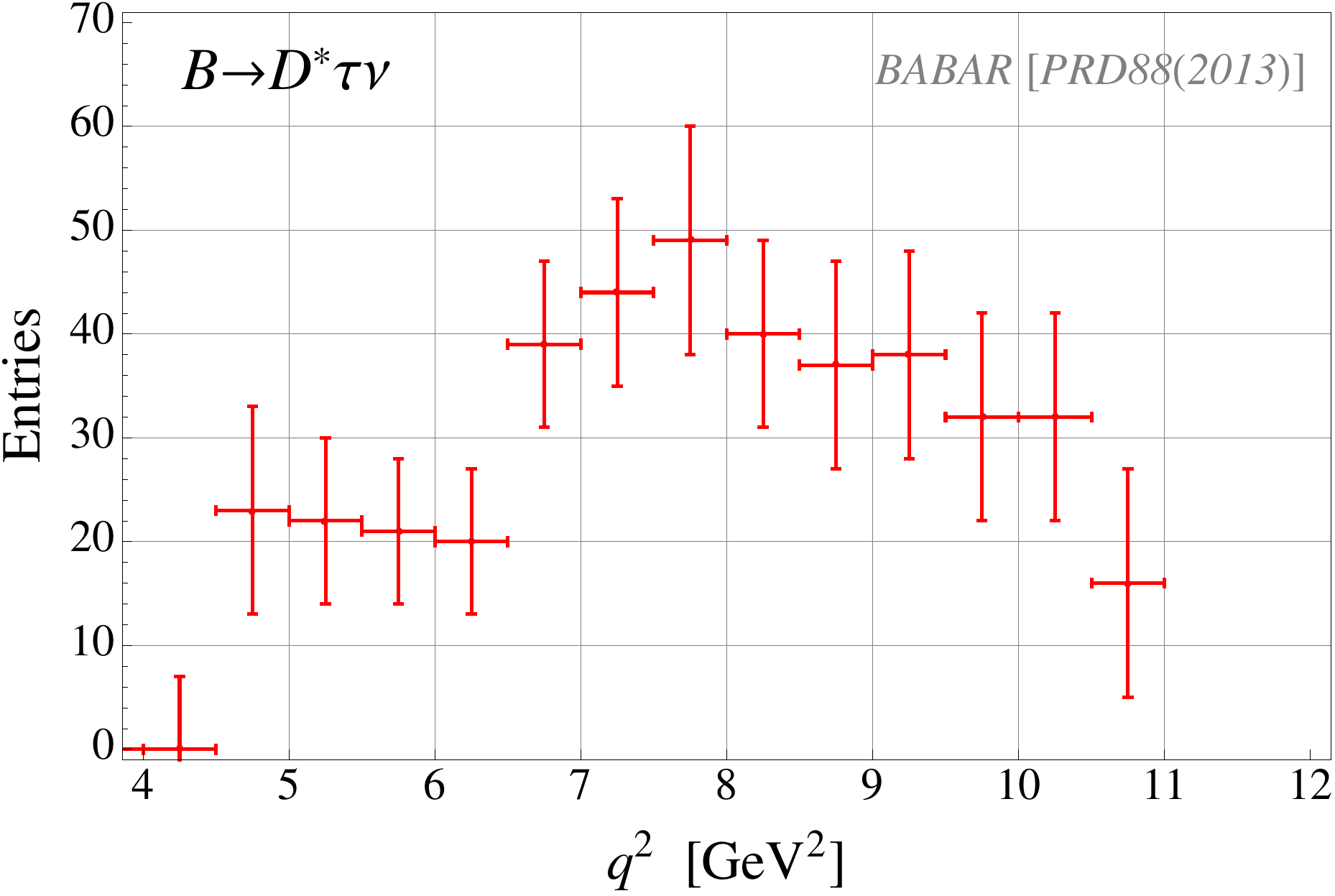}
   \caption{\footnotesize The measured background subtracted $q^2$ distributions for $\Bbar\to D\tau\nubar$ and $\Bbar\to \Dst\tau\nubar$ events, extracted from the \Babar~ data \protect\cite{Lees:2013uzd}.}
   \label{fig:BABAR_data}
\end{figure}

\begin{figure}[t!]\centering
   \includegraphics[width=0.45\linewidth]{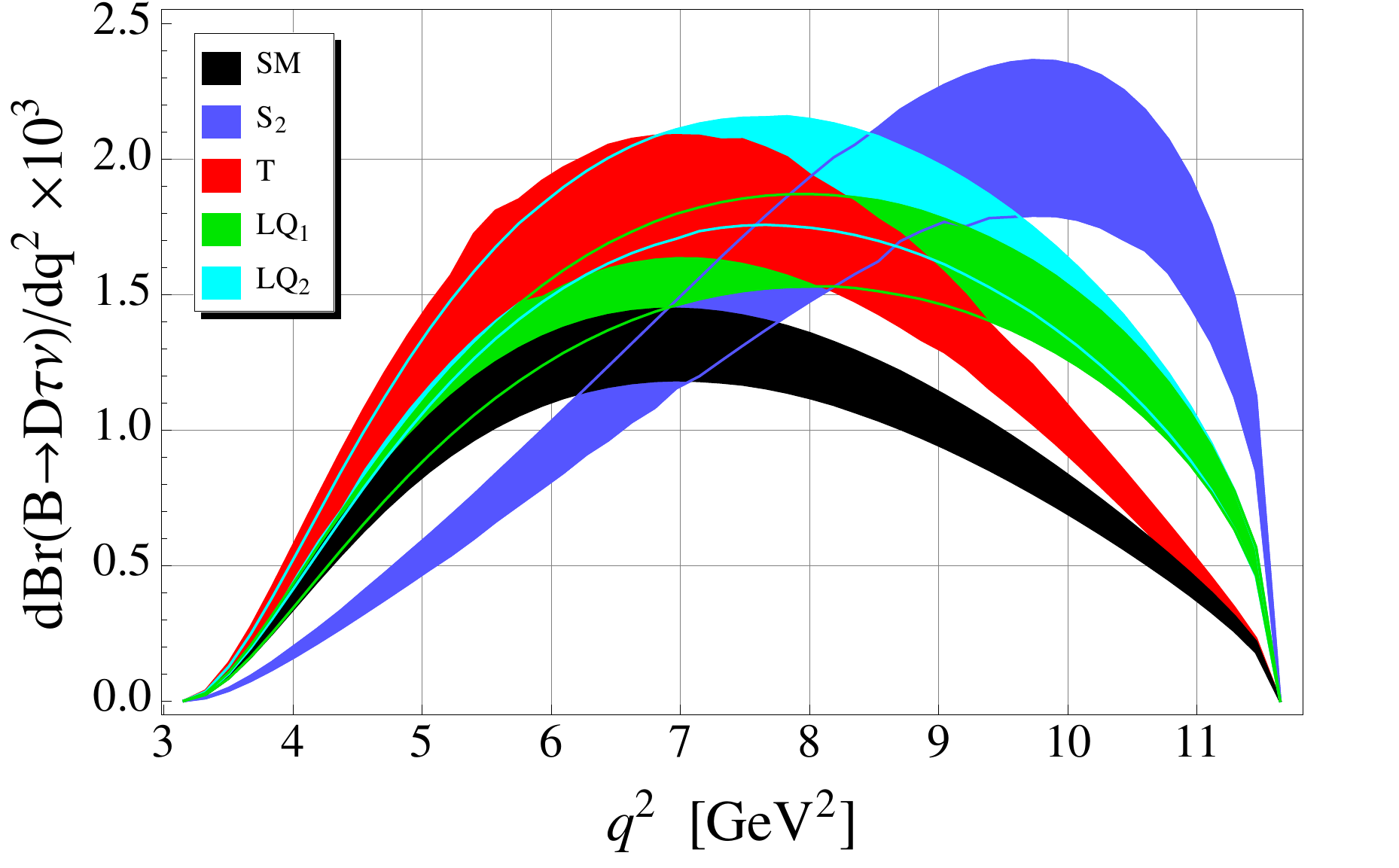} \hspace{3mm}
   \includegraphics[width=0.45\linewidth]{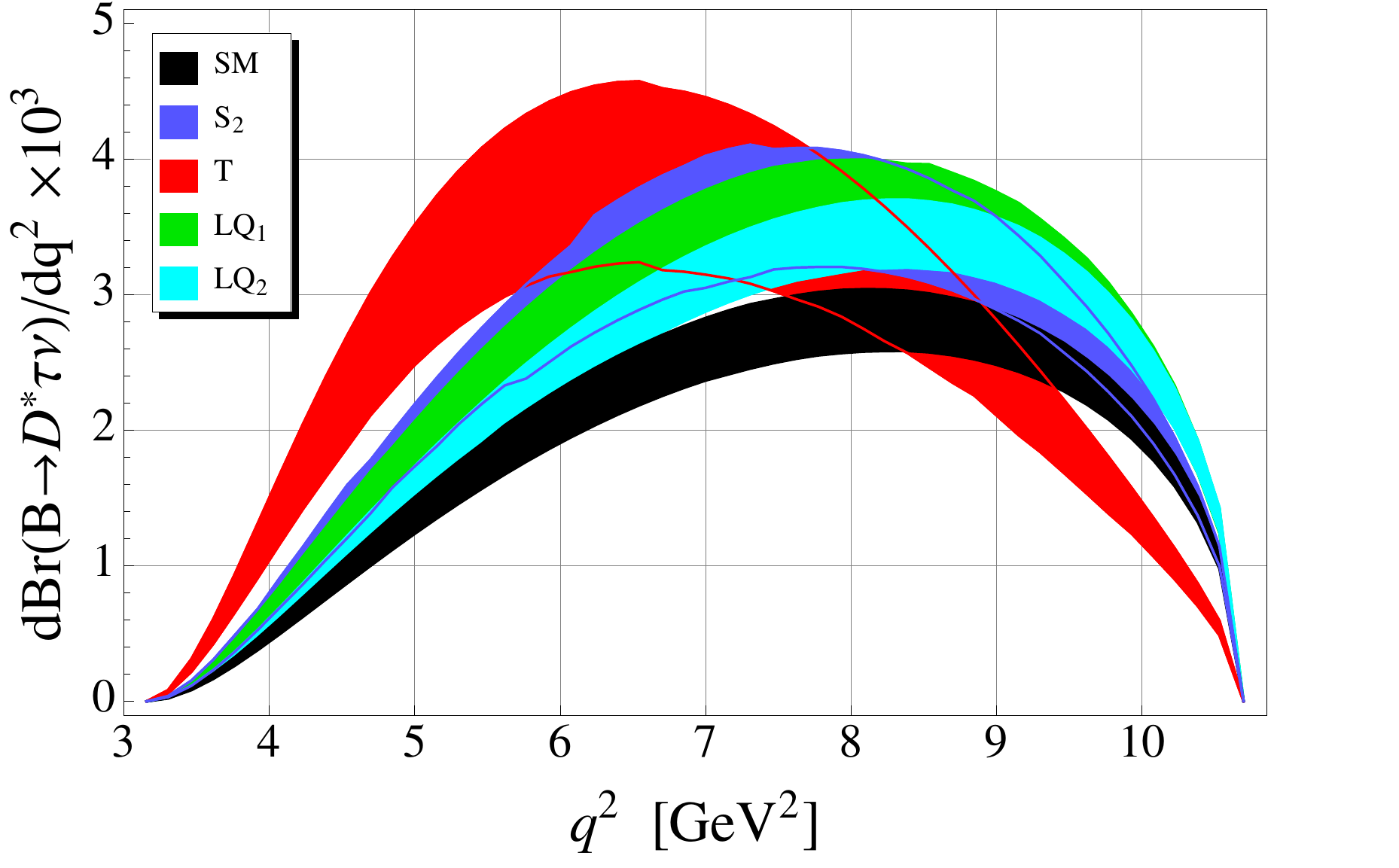}
   \caption{\footnotesize The differential branching fractions, predicted in the SM (black) and various NP scenarios listed in Section \ref{sec:Heff} : $S_2$ (blue), $T$ (red), LQ$_1$ (green) and LQ$_2$ (cyan). The width of each curve is due to the theoretical errors in the hadronic form factor parameters and the uncertainty of $V_{cb}$.}
   \label{fig:dBrdq2}
\end{figure}

To estimate the (dis)agreement between the measured and expected $q^2$ spectra, we extract the experimental numbers of signal events from Fig.~23 in Ref.~\cite{Lees:2013uzd} and compare them with the expectations of different scenarios listed in the previous section. We present the extracted experimental data points in Fig.~\ref{fig:BABAR_data}. In our study, we merge two last bins in Fig.~\ref{fig:BABAR_data} in order to satisfy the physical condition $q^2\leq(m_B-m_\DDst)^2$ and add corresponding errors in quadratures. The corresponding theoretical predictions for $d\B/dq^2$ distributions are presented in Fig.~\ref{fig:dBrdq2}. The width of each curve is due to the theoretical errors in the hadronic form factor parameters and the uncertainty of $V_{cb}=(41.1\pm1.3)\times10^{-3}$ \cite{Agashe:2014kda}.

Due to the lack of knowledge about the overall normalization of the spectra, in our study {\it we test only the shape of the distributions} and leave the normalization of the data to be a free parameter of each fit. This implies that the total efficiency is assumed to be a free parameter, constant for all $q^2$ bins and dependent on the tested model. The results on $p$ values are presented in Table \ref{tab:Babar_fit}. One can see from the table that the scalar (tensor) operator is disfavored by the observed $q^2$ distribution of the $\Bbar\to \DDst\tau\nubar$ decays.

\begin{table}[t!]
   {\scriptsize
   \begin{center}
      \begin{tabular}{|c|c|c|c|}
         \hline
         model & $\Bbar \to D\tau\nubar$ & $\Bbar \to \Dst\tau\nubar$ & $\Bbar \to (D+\Dst)\tau\nubar$ \\
         \hline
         SM & 54\% & 65\% & 67\% \\
         \hline
         $V_1$ & 54\% & 65\% & 67\% \\
         \hline
         $V_2$ & 54\% & 65\% & 67\% \\
         \hline
         $S_2$ & 0.02\% & 37\% & 0.1\% \\
         \hline
         $T$ & 58\% & 0.1\% & 1.0\% \\
         \hline
         LQ$_1$ & 13\% & 58\% & 25\% \\
         \hline
         LQ$_2$ & 21\% & 72\% & 42\% \\
         \hline
      \end{tabular}
   \end{center}
   }
   \caption{\footnotesize $p$ values for the fit of the \Babar~ data of $d\B/dq^2$ with various models.}
   \label{tab:Babar_fit}
\end{table}

In order to get rid of the dependence on $V_{cb}$, reduce theoretical uncertainties of hadronic form factors and increase the sensitivity of the $q^2$ dependencies to NP, we introduce the following quantities\,\footnote{The NP effects in $q^2$ distributions are also studied in Ref.~\cite{Duraisamy:2013pia,*Duraisamy:2014sna}.} :
\begin{equation}
   \begin{split}
      R_D(q^2) \equiv& {d\B(\Bbar\to D\tau\nubar)/dq^2 \over d\B(\Bbar\to D\ell\nubar)/dq^2} \, {\lambda_D(q^2) \over (m_B^2-m_D^2)^2} \, \left(1-{m_\tau^2 \over q^2}\right)^{-2} \,, \\
      R_\Dst(q^2) \equiv& {d\B(\Bbar\to\Dst\tau\nubar)/dq^2 \over d\B(\Bbar\to\Dst\ell\nubar)/dq^2} \, \left(1-{m_\tau^2 \over q^2}\right)^{-2} \,.
   \end{split}
   \label{eq:Rq2}
\end{equation}
Here for our convenience, to remove zero\,\footnote{In the SM, $d\B(\Bbar\to D\ell\nubar)/dq^2 \propto (H_V^{s})^2 \propto \lambda_D(q^2) \to 0$ for $q^2 \to q_{\rm max}^2$.} of $d\B(\Bbar\to D\ell\nubar)/dq^2$ at $q_{\rm max}^2=(m_B-m_D)^2$ and the phase space suppression of $d\B(B\to\DDst\tau\nubar)/dq^2$ at $q_{\rm min}^2 = m_\tau^2$, we introduced additional purely kinematic factors above.

\begin{figure}[t!]\centering
   \includegraphics[width=0.45\linewidth]{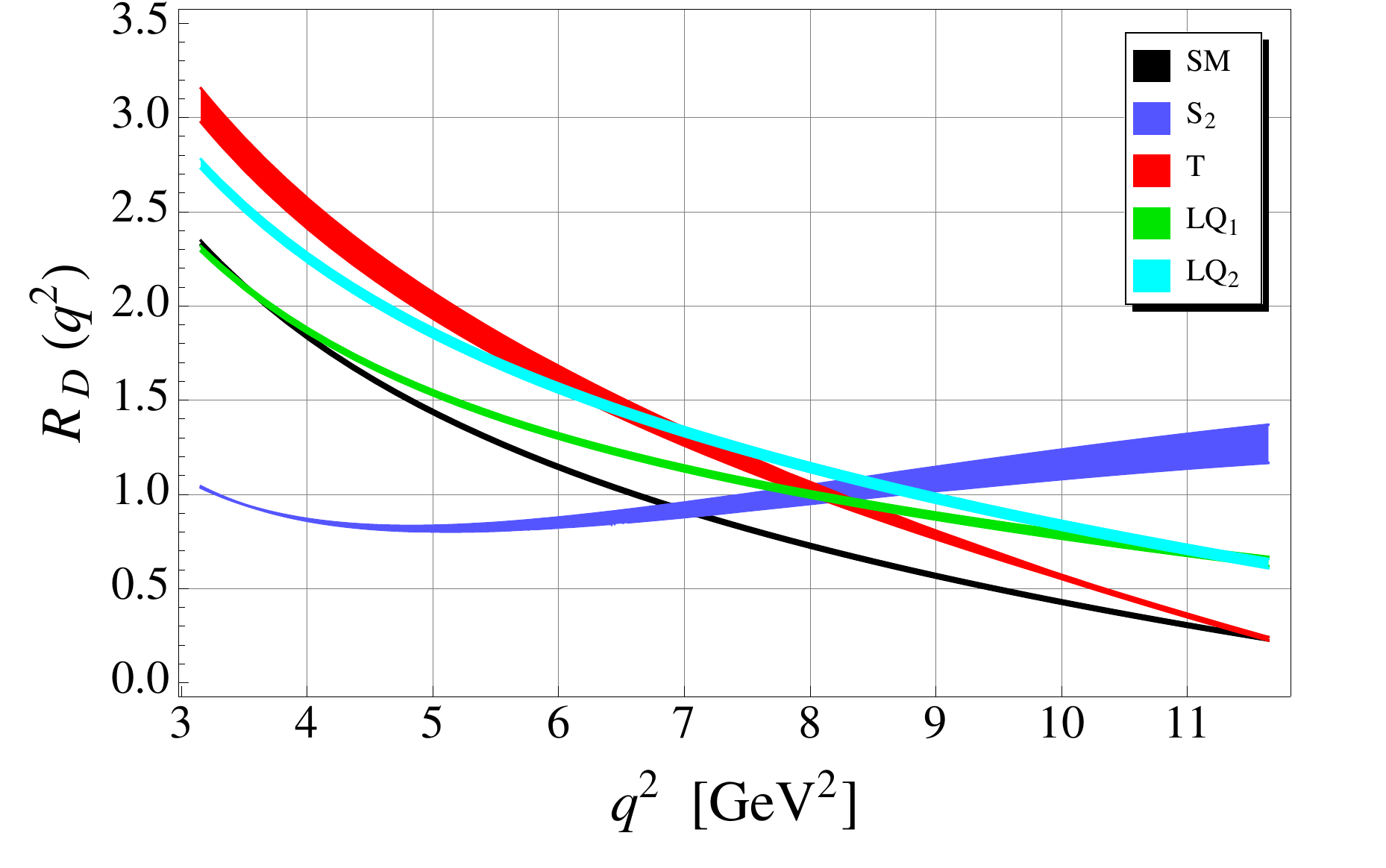} \hspace{3mm}
   \includegraphics[width=0.45\linewidth]{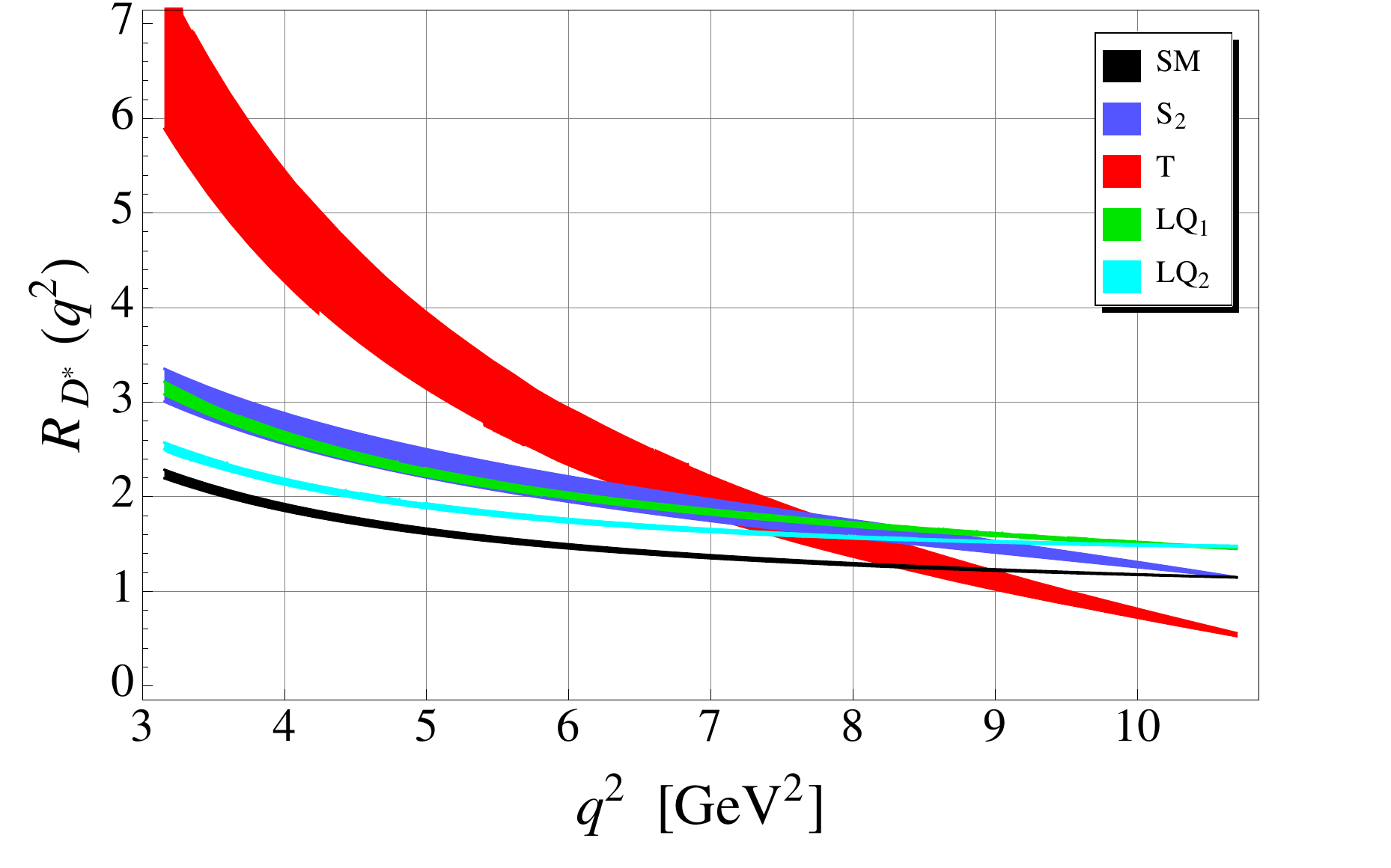}
   \caption{\footnotesize The $R_\DDst(q^2)$ distributions, predicted in the SM (black) and various NP scenarios listed in Section \ref{sec:Heff} : $S_2$ (blue), $T$ (red), LQ$_1$ (green) and LQ$_2$ (cyan). The width of each curve is due to the theoretical errors in the hadronic form factor parameters}
   \label{fig:Rq2}
\end{figure}

In Fig.~\ref{fig:Rq2}, for illustration, we show the $R_\DDst(q^2)$ distributions, predicted for the five scenarios described in Section \ref{sec:Heff}. The width of each curve is due to the theoretical errors in the hadronic form factor parameters, which are varied within $\pm1\sigma$ ranges. The distributions for the vector $V_{1,2}$ NP scenarios (with best fitted values of Wilson coefficients $C_{V_1}=0.16$ and $C_{V_2}=0.01\pm0.60i$ respectively) have small theoretical uncertainties as in the SM, but are practically indistinguishable from the distribution of the tensor (LQ$_1$) NP scenario for the $D$($\Dst$) mode. Therefore we omit plotting them in Fig.~\ref{fig:Rq2}.

 We find that $R_D(q^2)$ is very sensitive to the scalar contribution and $R_\Dst(q^2)$ is more sensitive to the tensor operator. Moreover, one can easily see from Figs. \ref{fig:dBrdq2} and \ref{fig:Rq2} that the theoretical uncertainties in $R_\DDst(q^2)$ are significantly smaller than those  of the differential branching fractions. Hence, the $R_\DDst(q^2)$ distributions provide a good test of NP in addition to $R(\DDst)$.

\section{Discriminative potential at Belle II}\label{sec:BelleII}

In order to demonstrate the discriminating power of $R_\DDst(q^2)$, we simulate ``experimental data'' for the binned $R_\DDst(q^2)$ distributions, assuming one of the scenarios, listed in Section \ref{sec:Heff}, that can explain the observed deviation in $R(D)$ and $R(\Dst)$, and compare them with other various model predictions by calculating $\chi^2$ defined in the following way:
\begin{equation}
   \chi^2 = \sum_{i,j=1}^{N_{\rm bins}}
   ( R_i^{\rm exp} - R_i^{\rm model} )
   ( V^{\rm exp} + V^{\rm model} )_{ij}^{-1}
   ( R_j^{\rm exp} - R_j^{\rm model} ) \,,
\end{equation}
where $i$ and $j$ denote the $q^2$-bin indices, $V^{\rm exp}$ and $V^{\rm model}$ are the experimental and theoretical covariance matrices of the simulated ``experimental data'' and the tested model respectively. Here the binned $R_i$ is defined as $R_i=(N_i^\tau/N_i^\ell)f(q_i^2)$ with $f(q_i^2)$ for shortness denoting purely kinematic factors introduced in Eq.~\eqref{eq:Rq2}, where $N_i^{\tau,\ell}$ are the numbers of signal events in the $i$\textsuperscript{th} bin for a given luminosity. We evaluate $N_i^{\tau,\ell}$ for each benchmark scenario using the central values of the hadronic parameters.

For model predictions, the uncertainties of the HQET hadronic form factors and the quark masses are taken into account in the calculation of $V^{\rm model}$, defined as
\begin{equation}
   V_{ij}^{\rm model} = \langle ( R_i^{\rm model} - \langle R_i^{\rm model} \rangle ) ( R_j^{\rm model} - \langle R_j^{\rm model} \rangle ) \rangle \,.
   \label{eq:Vth}
\end{equation}
The HQET form factor parameters are assumed to have the Gaussian distribution, while $m_b\pm m_c$ are varied uniformly in the corresponding $\pm1\sigma$ ranges, $m_b-m_c=(3.45\pm0.05)~\GeV$ and $m_b+m_c=(6.2\pm0.4)~\GeV$.

Due to the lack of the detailed detector and background simulation, we simply assume that ($i$)~$V^{\rm exp}$ is diagonal, ($ii$)~systematic errors are of the same as statistical ones, ($iii$) to be more conservative, we add systematic and statistical errors linearly. Accordingly, the covariance matrix of the ``experimental data'' is evaluated as
\begin{equation}
   V_{ij}^{\rm exp} \approx \delta_{ij}(2\delta_{\rm stat}R_i^{\rm exp})^2 \,.
\end{equation}

Neglecting the error of the number of signal events in each bin for the $e,\mu$ modes $\delta N_i^\ell$ (compared to the $\tau$ mode) due to the large expected statistics at SuperKEKB/Belle~II, we estimate $\delta_{\rm stat} R_i^{\rm exp}$ as follows:
\begin{equation}
   \delta_{\rm stat} R_i^{\rm exp} \approx {\delta N_i^\tau \over N_i^\ell}{\varepsilon_i^\ell \over \varepsilon_i^\tau} f(q_i^2) \approx {1 \over \sqrt{N_{B\Bbar}\,\varepsilon_i^\tau}}{\sqrt{\B_i^\tau} \over \B_i^\ell} f(q_i^2) \,,
\end{equation}
where $N_{B\Bbar}=\L\times\sigma(e^+e^-\to B\Bbar)$ is the number of produced $B\Bbar$ pairs for an integrated luminosity $\L$, $\B_i^{\tau,\ell}$ denote the branching fractions integrated over the $i$\textsuperscript{th} bin. For simplicity, taking the efficiency $\varepsilon_i^\tau$ to be constant for all bins, we estimate it to be $\varepsilon_i^\tau \approx \varepsilon_{\rm tot}^\tau \simeq 10^{-4}$, using the $\Babar$ result on total number of signal events. 

In Table~\ref{tab:lumi} we present our results on luminosities for various sets of simulated ``data'' and a tested model, required to exclude the model at 99.9\% C.L. using binned $R_D(q^2)$ and $R_\Dst(q^2)$ distributions. In parentheses, for comparison, we present the required luminosity using the $R(D)$ and $R(\Dst)$ ratios. The cross mark means that it's impossible to distinguish ``data'' and model at 99.9\% C.L. due to very small $\chi^2$ values (however, the discrimination at 68\% C.L. is still possible for some models). This occurs in the cases when statistical errors vanish ($\L\to\infty$) and theoretical uncertainties remain non-negligible. As one can see from the table, some cases of ``data''-model (e.g. $S_2$-$T$ or $S_2$-$V_{1,2}$) can be already tested using the \Babar~ and Belle statistics ($\L_{\scriptsize{\Babar}}=426$~fb$^{-1}$, $\L_{\rm Belle}=711$~fb$^{-1}$). In order to test the leptoquark scenarios, one needs about 1-6~ab$^{-1}$, which will be achieved at the early stage of the Belle II experiment. To discriminate the $V_1$ and $V_2$ NP scenarios turns out to be practically impossible due to too high required luminosity that cannot be achieved at near future colliders.

To find out which of two methods, using $R_\DDst(q^2)$ or $R(\DDst)$, is more effective (i.e. requires a smaller luminosity) and more sensitive to a particular NP scenario, we illustrate the results of Table~\ref{tab:lumi} in a simple way in Table~\ref{tab:Rq2_vs_R}. Small circles and squares represent the advantage of $R_\DDst(q^2)$ and $R(\DDst)$ respectively. Double circles correspond to the case when only $R_\DDst(q^2)$ is effective. Cross marks denote the impossibility of discrimination by either of the two methods. As for the SM, we do not need $R_\DDst(q^2)$ since the present experimental data of $R(\DDst)$ have already shown the significant deviation from the SM as explained in Section \ref{sec:intro}. 

As can been seen from Table~\ref{tab:Rq2_vs_R}, for the ``data''-model cases LQ$_2(V_{1,2})$-$V_{1,2}$(LQ$_2$) and LQ$_{2(1)}$-LQ$_{1(2)}$, $R(\DDst)$ turn out to be more advantageous quantities to be studied. On the other hand, if we assume ``data'' to be e.g. $S_2$ or $T$, the binned $q^2$ distributions become more profitable for discrimination of other NP models. Moreover, only $R_\DDst(q^2)$ can clearly distinguish the $S_2$-$T$ and $T$-$S_2$ cases. To summarise, among the 36 cases listed in Table~\ref{tab:Rq2_vs_R}, in 22 cases the study of $q^2$ distributions turns out to be more advantageous and has a lower luminosity cost, and in 15 cases only $R_\DDst(q^2)$ can discriminate ``data'' and models at 99.9\% C.L.

\begin{table}[t!]
   {\scriptsize
   \begin{center}
      \begin{tabular}{|cc||c|c|c|c|c|c|c|}
         \hline
         \multicolumn{2}{|c||}{\multirow{2}{*}{$\L~[{\rm fb}^{-1}]$}} & \multicolumn{7}{c|}{\bf model} \\
         \hhline{|~~||-------}
         & & SM & $V_1$ & $V_2$ & $S_2$ & $T$ & LQ$_1$ & LQ$_2$ \\
%         \hhline{|==||=|=|=|=|=|=|=|}
         \hhline{|--||-|-|-|-|-|-|-|}
         \multicolumn{1}{|c|}{} & $V_1$ & \begin{tabular}{c} 1170 \\ (270) \end{tabular} &  & \begin{tabular}{c} $10^6$ \\ (\ding{53}) \end{tabular} & \begin{tabular}{c} 500 \\ (\ding{53}) \end{tabular} & \begin{tabular}{c} 900 \\ (\ding{53}) \end{tabular} & \begin{tabular}{c} 4140 \\ (\ding{53}) \end{tabular} & \begin{tabular}{c} 2860 \\ (1390) \end{tabular} \\
         \hhline{~--------}
         \multicolumn{1}{|c|}{} & $V_2$ & \begin{tabular}{c} 1140 \\ (270) \end{tabular} & \begin{tabular}{c} $10^6$ \\ (\ding{53}) \end{tabular} &  & \begin{tabular}{c} 510 \\ (\ding{53}) \end{tabular} & \begin{tabular}{c} 910 \\ (\ding{53}) \end{tabular} & \begin{tabular}{c} 4210 \\ (\ding{53}) \end{tabular} & \begin{tabular}{c} 3370 \\ (1960) \end{tabular} \\
         \hhline{~--------}
         \multicolumn{1}{|c|}{\parbox[t]{2mm}{\multirow{2}{*}{\rotatebox[origin=c]{90}{\bf ``data''}}}} & $S_2$ & \begin{tabular}{c} 560 \\ (290) \end{tabular} & \begin{tabular}{c} 560 \\ (13750) \end{tabular} & \begin{tabular}{c} 540 \\ (36450) \end{tabular} &  & \begin{tabular}{c} 380 \\ (\ding{53}) \end{tabular} & \begin{tabular}{c} 1310 \\ (35720) \end{tabular} & \begin{tabular}{c} 730 \\ (4720) \end{tabular} \\
         \hhline{~--------}
         \multicolumn{1}{|c|}{} & $T$ & \begin{tabular}{c} 600 \\ (270) \end{tabular} & \begin{tabular}{c} 680 \\ (\ding{53}) \end{tabular} & \begin{tabular}{c} 700 \\ (\ding{53}) \end{tabular} & \begin{tabular}{c} 320 \\ (\ding{53}) \end{tabular} &  & \begin{tabular}{c} 620 \\ (\ding{53}) \end{tabular} & \begin{tabular}{c} 550 \\ (1980) \end{tabular} \\
         \hhline{~--------}
         \multicolumn{1}{|c|}{} & LQ$_1$ & \begin{tabular}{c} 1010 \\ (270) \end{tabular} & \begin{tabular}{c} 4820 \\ (\ding{53}) \end{tabular} & \begin{tabular}{c} 4650 \\ (\ding{53}) \end{tabular} & \begin{tabular}{c} 1510 \\ (\ding{53}) \end{tabular} & \begin{tabular}{c} 800 \\ (\ding{53}) \end{tabular} &  & \begin{tabular}{c} 5920 \\ (1940) \end{tabular} \\
         \hhline{~--------}
         \multicolumn{1}{|c|}{} & LQ$_2$ & \begin{tabular}{c} 1020 \\ (250) \end{tabular} & \begin{tabular}{c} 3420 \\ (1320) \end{tabular} & \begin{tabular}{c} 3990 \\ (1820) \end{tabular} & \begin{tabular}{c} 1040 \\ (20560) \end{tabular} & \begin{tabular}{c} 650 \\ (4110) \end{tabular} & \begin{tabular}{c} 5930 \\ (1860) \end{tabular} &  \\
         \hline
      \end{tabular}
   \end{center}
   }
   \caption{\footnotesize Luminosity required to discriminate various simulated ``data'' and tested model sets at 99.9\%~C.L. using $R_\DDst(q^2)$ or $R(\DDst)$ (in parentheses).}
   \label{tab:lumi}
\end{table}

\begin{table}[t!]
   {\scriptsize
   \begin{center}
      \begin{tabular}{|cc||c|c|c|c|c|c|c|}
         \hline
         \multicolumn{2}{|c||}{\multirow{2}{*}{}} & \multicolumn{7}{c|}{\bf model} \\
         \hhline{|~~||-------}
         & & SM & $V_1$ & $V_2$ & $S_2$ & $T$ & LQ$_1$ & LQ$_2$ \\
%         \hhline{|==||=|=|=|=|=|=|=|}
         \hhline{|--||-|-|-|-|-|-|-|}
         \multicolumn{1}{|c|}{} & $V_1$ & $\markR$ &  & \ding{56} & $\MarkRq$ & $\MarkRq$ & $\MarkRq$ & $\markR$ \\
         \hhline{~--------}
         \multicolumn{1}{|c|}{} & $V_2$ & $\markR$ & \ding{56} &  & $\MarkRq$ & $\MarkRq$ & $\MarkRq$ & $\markR$ \\
         \hhline{~--------}
         \multicolumn{1}{|c|}{\parbox[t]{2mm}{\multirow{2}{*}{\rotatebox[origin=c]{90}{\bf ``data''}}}} & $S_2$ & $\markR$ & $\markRq$ & $\markRq$ &  & $\MarkRq$ & $\markRq$ & $\markRq$ \\
         \hhline{~--------}
         \multicolumn{1}{|c|}{} & $T$ & $\markR$ & $\MarkRq$ & $\MarkRq$ & $\MarkRq$ &  & $\MarkRq$ & $\markRq$ \\
         \hhline{~--------}
         \multicolumn{1}{|c|}{} & LQ$_1$ & $\markR$ & $\MarkRq$ & $\MarkRq$ & $\MarkRq$ & $\MarkRq$ &  & $\markR$ \\
         \hhline{~--------}
         \multicolumn{1}{|c|}{} & LQ$_2$ & $\markR$ & $\markR$ & $\markR$ & $\markRq$ & $\markRq$ & $\markR$ &  \\
         \hline
      \end{tabular}
   \end{center}
   }
   \caption{\footnotesize Comparison of two discrimination methods, using $R_\DDst(q^2)$ (circle) or $R(\DDst)$ (square): the method requiring a smaller luminosity to distinguish ``data'' and theoretical model at 99.9\% C.L. is more advantageous. Double circle corresponds to the case when only $R_\DDst(q^2)$ is effective and can distinguish scenarios. Cross marks denote the impossibility of discrimination by either of the two methods.}
   \label{tab:Rq2_vs_R}
\end{table}

\begin{figure}[p!]\centering
   \includegraphics[width=0.32\linewidth]{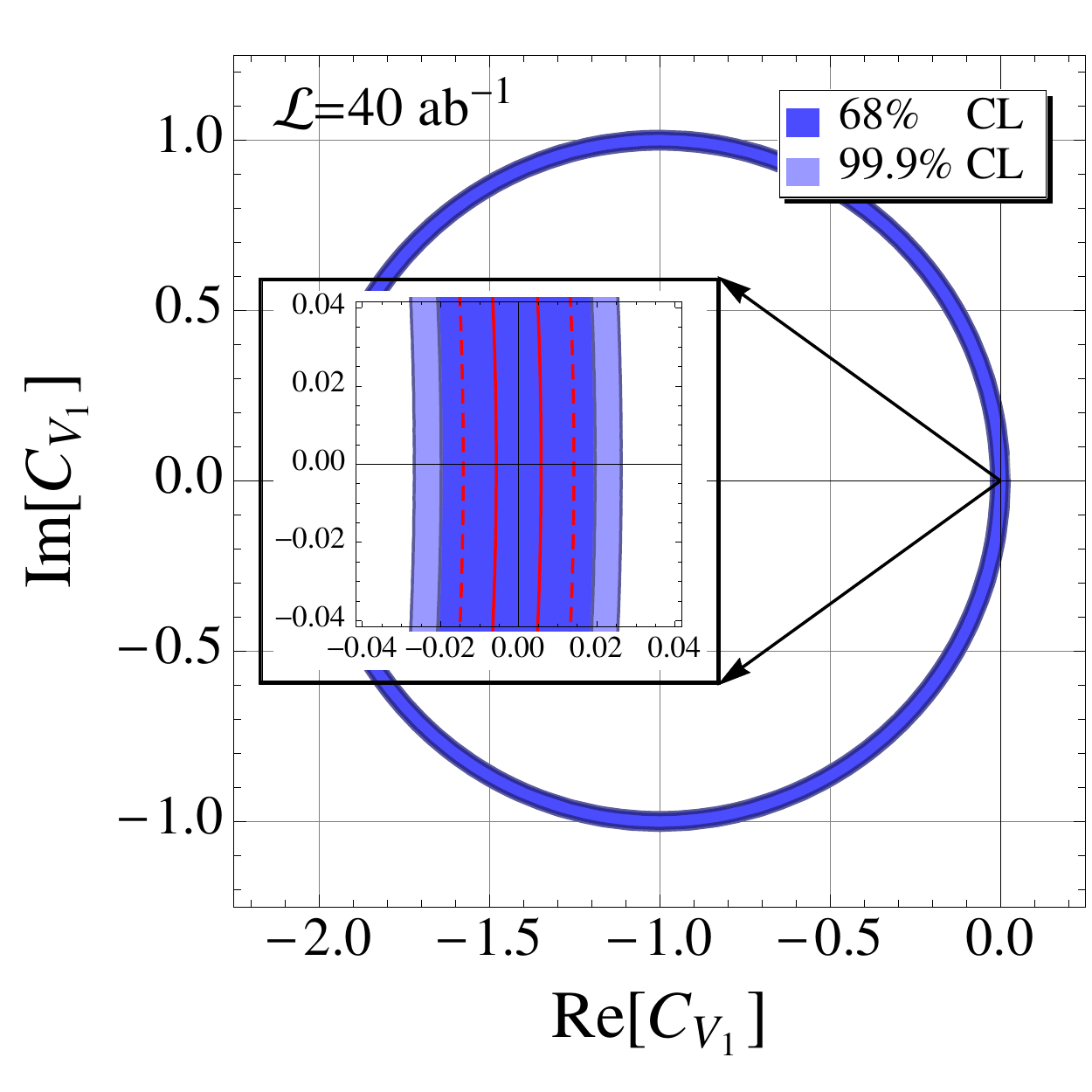}
   \includegraphics[width=0.32\linewidth]{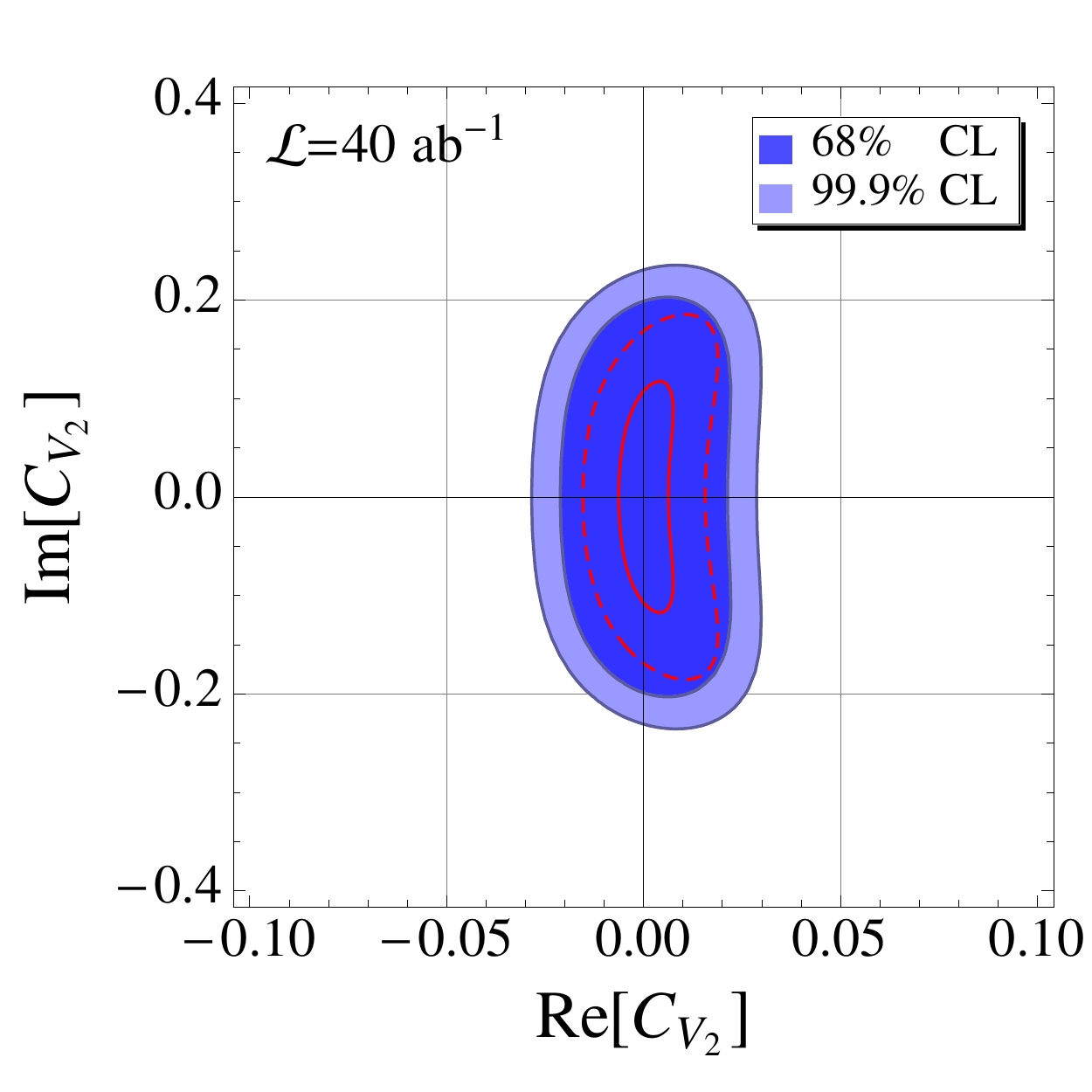}
   \includegraphics[width=0.32\linewidth]{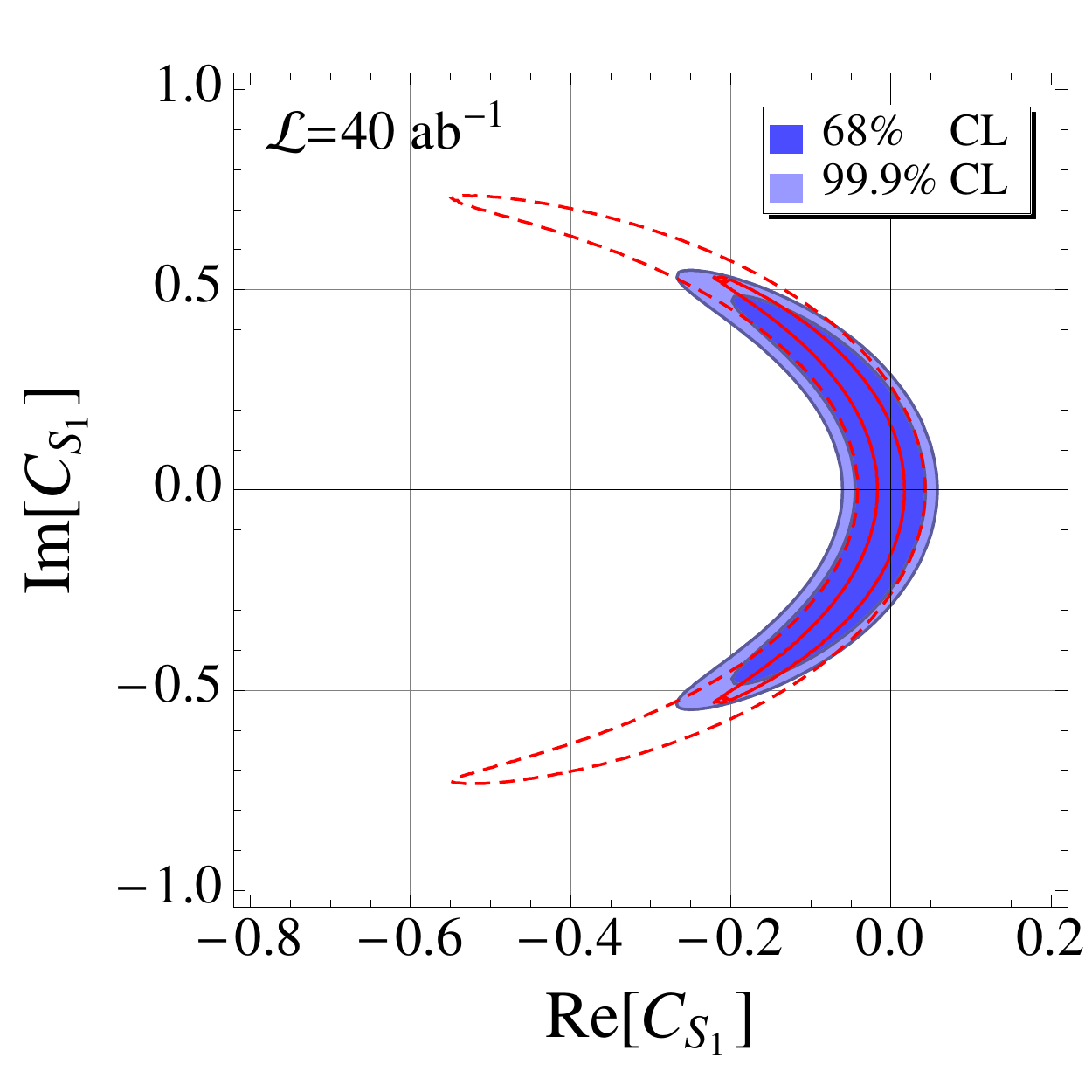}

   \includegraphics[width=0.32\linewidth]{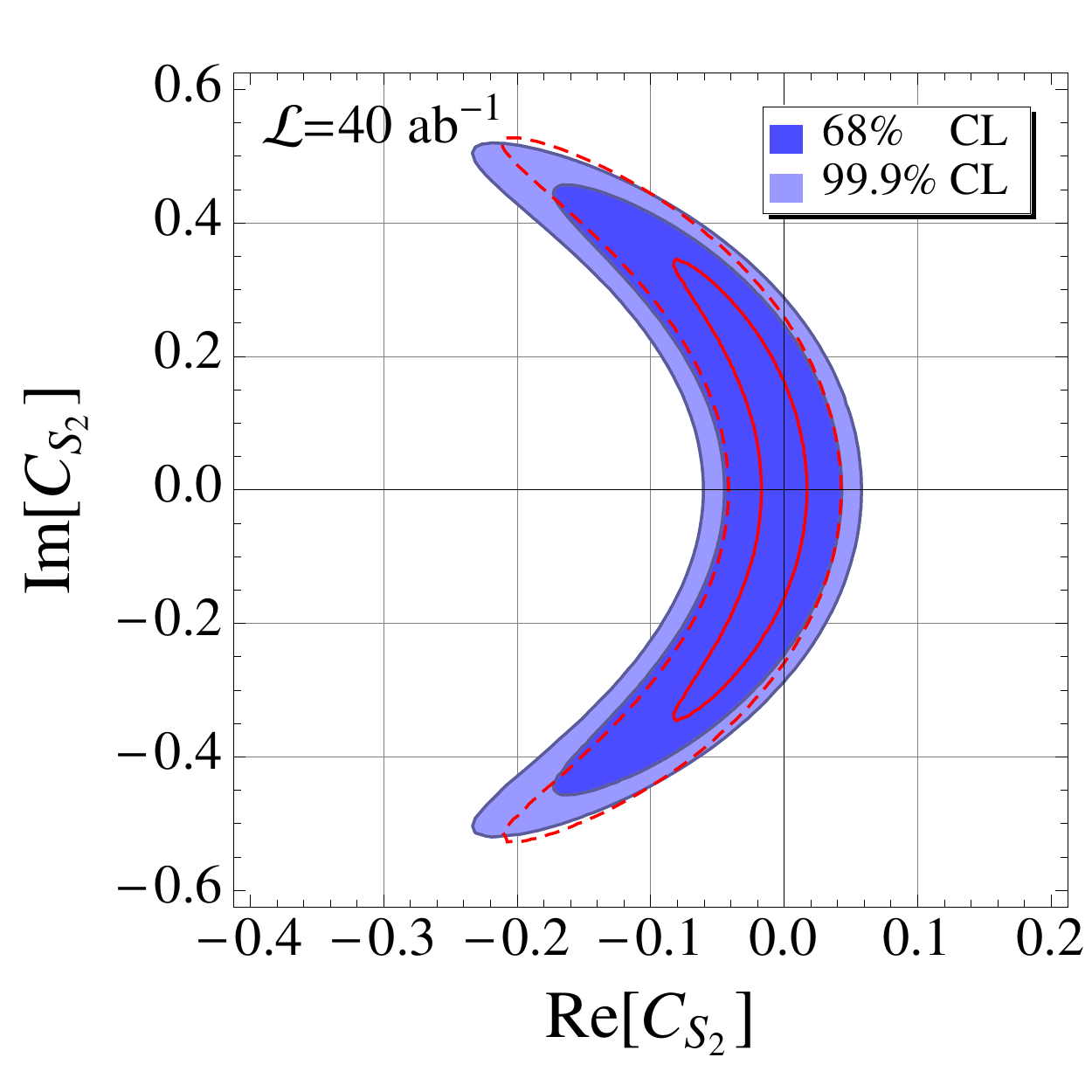}
   \includegraphics[width=0.32\linewidth]{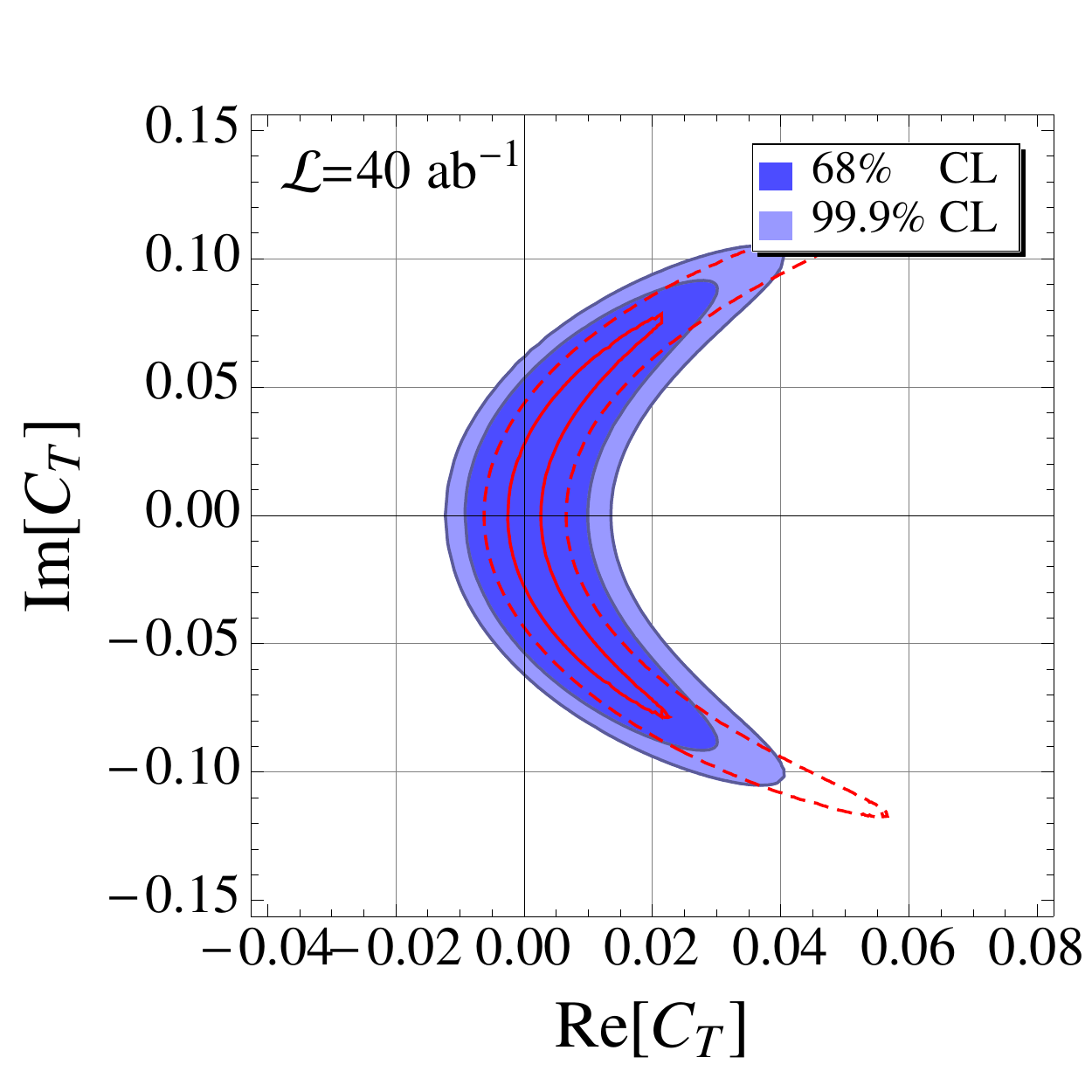}
   \includegraphics[width=0.32\linewidth]{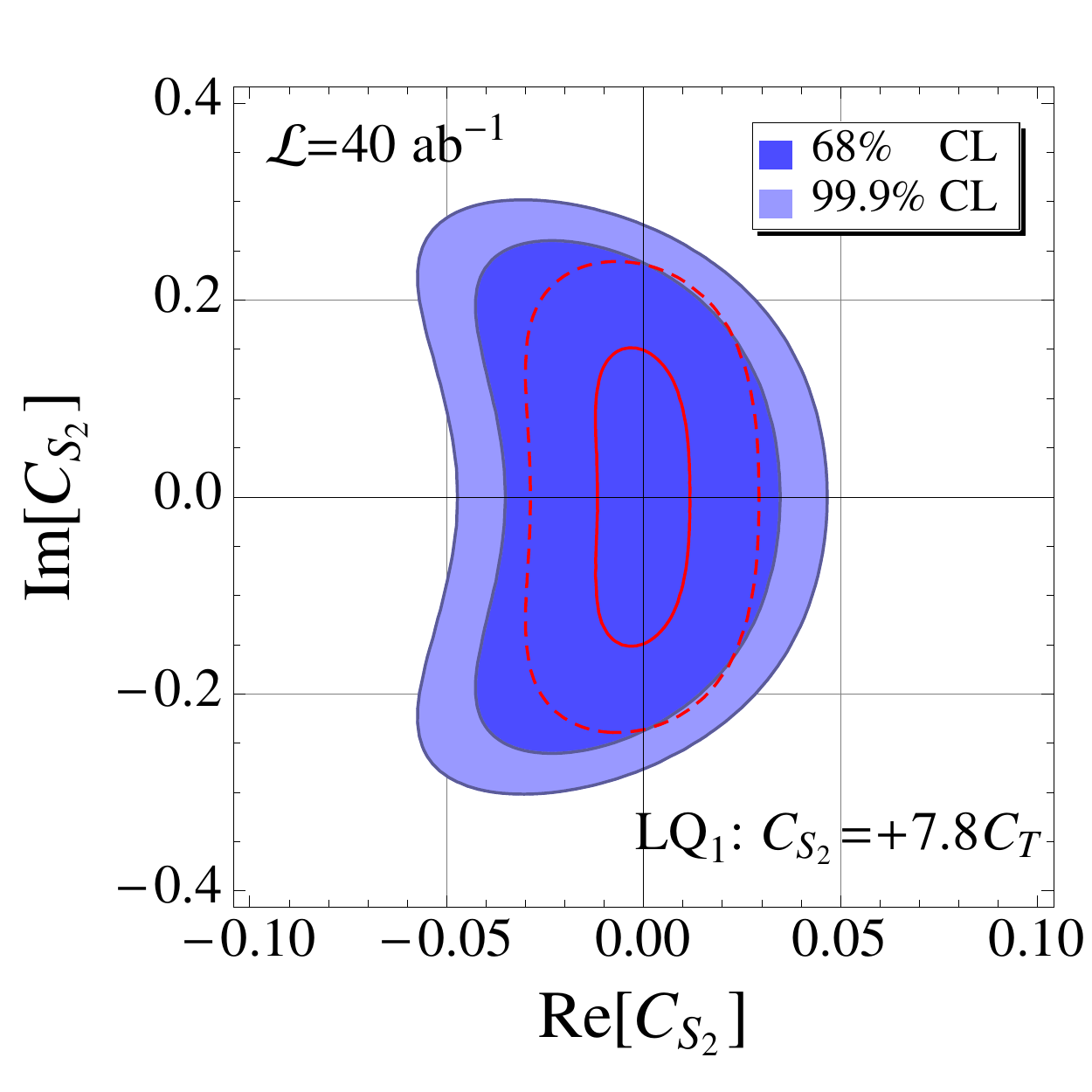}

   \includegraphics[width=0.32\linewidth]{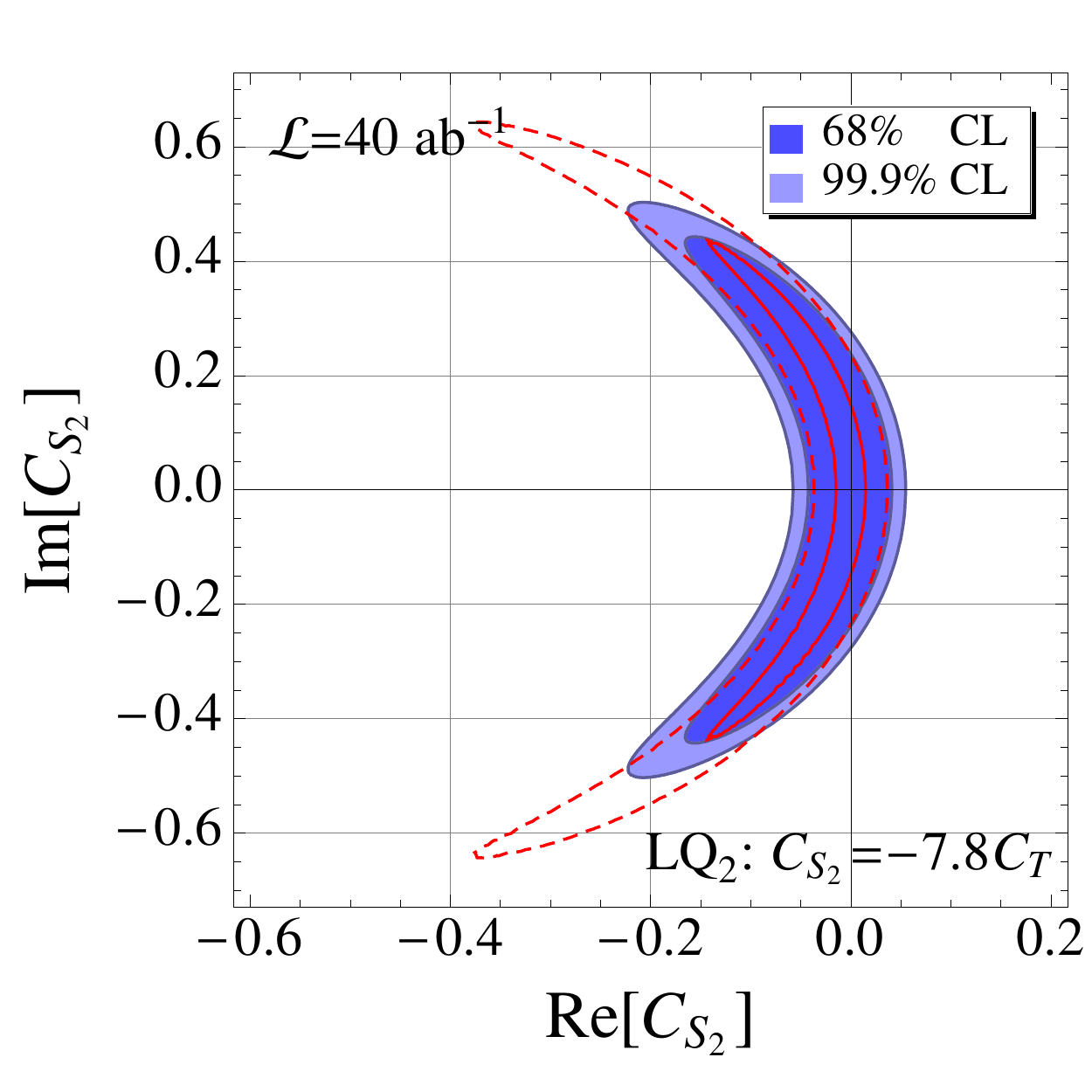}
   \caption{\footnotesize Constraints on the Wilson coefficients at the $m_b$ scale. The constraints are obtained from the $\chi^2$ fit of binned $R_D(q^2)$ and $R_\Dst(q^2)$ assuming the future experimental measurements at Belle~II for the integrated luminosity 40~ab$^{-1}$ to be perfectly consistent with the SM predictions. The red solid(dashed) lines correspond to the constraints at 68\% (99.9\%) C.L. coming from the $q^2$-integrated $R(\DDst)$.}
   \label{pic:CX_BelleII}
\end{figure}

To clarify the sensitivity to NP Wilson coefficients in the Belle II experiment, in Fig.~\ref{pic:CX_BelleII} we present constraints on the Wilson coefficients, obtained from the $\chi^2$ fit of binned $R_D(q^2)$ and $R_\Dst(q^2)$ for the integrated luminosity of 40~ab$^{-1}$, assuming the ``data'' to be perfectly consistent with the SM predictions. The dark (light) blue regions represent the expected 68\% (99.9\%)~C.L. constraints from $R_D(q^2)$ and $R_\Dst(q^2)$. For comparison, we show the 68\% (99.9\%)~C.L. allowed regions, represented by red solid (dashed) lines, from $R(D)$ and $R(\Dst)$. Due to the large statistics of the $\Bbar \to \DDst\ell\nubar_\ell$ events at the Belle~II experiment, it will be possible to improve significantly the precision of the HQET form factor parameters. Therefore, making Fig.~\ref{pic:CX_BelleII}, we suppose that the overall theoretical uncertainties of $R(\DDst)$ and $R_\DDst(q^2)$ will be reduced by factor 2, i.e. the covariance matrix in Eq.~\eqref{eq:Vth} is reduced by factor 4. We verified numerically that this approximation gives practically identical results to those obtained improving the accuracy of all HQET parameters and quark masses by factor 2.

Using the obtained constraints on the NP Wilson coefficients $C_{S_2}(m_b)$ and $C_T(m_b)$ in Fig.~\ref{pic:CX_BelleII} and performing the renormalization from $m_b$ to $M_{\rm NP}$ scale\,\footnote{The vector and axial vector currents are not renormalized because their anomalous dimensions vanish.}, one can study potential future constraints on NP couplings and masses. Here we assume for simplicity,
\begin{equation}
   C_X(M_{\rm NP}) \approx {1 \over 2\sqrt2 G_F V_{cb}} {\lambda\lambda^{\prime*} \over M_{\rm NP}^2} \,,
   \label{eq:CXNP}
\end{equation}
where $\lambda$ and $\lambda^\prime$ denote the general couplings of new heavy particles to quarks and leptons at the $M_{\rm NP}$ scale. Assuming the NP couplings $\lambda,\lambda^\prime \sim 1$, one can probe and constrain new particle masses as $M_{\rm NP} \gtrsim$ 5(7),  5(6), 7(10), 5(7), 5(6)~TeV for $V_{1,2}$, $S_{1,2}$, $T$, LQ$_1$ and LQ$_2$ NP types respectively, using the constraints from $R_\DDst(q^2)$ ($R(\DDst)$). Thus, one observes that if the experimental data is SM-like, $R(\DDst)$ turn out to be more advantageous observables to constrain NP scale, implying the statistical benefit of integrated quantities.

\section{Conclusions}

We studied NP effects in the $q^2$ distributions of the decay rates in $\Bbar \to D\tau\nubar$ and $\Bbar \to \Dst\tau\nubar$ considering the generic vector, scalar and tensor operators with Wilson coefficients of $O(1)$ that can describe the present experimental data quite well. We examined the currently available differential branching fractions of \Babar~ and estimated the $p$ values of the fit for various NP scenarios presented. We found that the scalar (tensor) operator is disfavored with $p=$0.1\% (1.0\%) by the observed differential branching fractions, however, their combinations that appear in leptoquark models are consistent with the data.

In order to cancel the dependence on $V_{cb}$ and reduce theoretical uncertainties, we introduced new quantities $R_\DDst(q^2)$ that turned out to be a very good tool for discriminating different NP scenarios in the future SuperKEKB/Belle~II experiment. 
In particular, $R_D(q^2)$ is very sensitive to the scalar contribution, and $R_\Dst(q^2)$ is more sensitive to the tensor operator. Hence, in addition to the $R(\DDst)$ determination, the study of $R_\DDst(q^2)$ distributions can can provide a good test of NP (including the leptoquark scenarios).

In order to evaluate the discriminating power of $R_\DDst(q^2)$, we simulated ``experimental data'' for the binned $R_\DDst(q^2)$ distributions, assuming one of the NP scenarios consistent with the observed deviation in $R(D)$ and $R(\Dst)$, and compared them with other theoretical model predictions.
We estimated luminosities required to exclude the tested models for various simulated ``data'' at 99.9\% C.L. using binned $R_\DDst(q^2)$ distributions as well as $R(\DDst)$.
It was found that over 36 possible scenarios listed in Table~\ref{tab:Rq2_vs_R}, in 22 cases studying $q^2$ distributions turned out to be more advantageous and have lower luminosity costs than $R(\DDst)$ measurement, and in 15 cases only $R_\DDst(q^2)$ can clearly discriminate ``data'' and models.
In addition, if the experimental data is SM-like, $R(\DDst)$ are more advantageous observables to constrain NP scale as reasonably understood by the statistical benefits of the integrated quantities.

Although in the future Belle II experiment statistical and systematic errors will be significantly reduced, theoretical uncertainties may remain non-negligible or even comparable to experimental ones. Therefore, for precise theoretical evaluation of $R(\DDst)$ and $R_\DDst(q^2)$ our knowledge of hadronic form factors (in particular, $1/m_{b,c}$ corrections) must be improved. In addition, to determine the scalar and tensor form factors we use equations of motion that involves the uncertainties related to the quark masses. Thus, new theoretical calculations using lattice QCD would be very helpful in future.

\section*{Acknowledgements}

This work is supported in part by JSPS KAKENHI Grant Numbers 25400257 (M.T.) and 2402804 (A.T.), and by IBS-R018-D1 (R.W.).

\vspace{1cm}

\bibliographystyle{utphys.bst}
\bibliography{bibliography}

\end{document}